\documentclass[12pt]{iopart}
\usepackage{iopams}  
\usepackage{setstack}  
\usepackage{verbatim} 
\usepackage{graphicx} 
\usepackage{bbold}  
\usepackage{esint}

\begin{document}
\title{A continuous Riemann-Hilbert problem for colliding plane gravitational waves}

\author{Stefan Palenta and Reinhard Meinel}

\address{Theoretisch-Physikalisches Institut, Friedrich-Schiller-Universit\"{a}t
Jena, Max-Wien-Platz 1, 07743 Jena, Germany}
\eads{\mailto{spalenta@gmail.com}, \mailto{meinel@tpi.uni-jena.de}}

\begin{abstract}
We present the foundations of a new solution technique for the characteristic initial value problem (IVP) of colliding plane gravitational waves. It has extensive similarities to the approach of Alekseev and Griffiths in 2001, but we use an inverse scattering method with a Riemann-Hilbert problem (RHP), which allows for a transformation to a continuous RHP with a solution given in terms of integral equations for non-singular functions. Ambiguities in this procedure lead to the construction of a family of spacetimes containing the solution to the IVP. Therefore the described technique also serves as an interesting solution generating method. The procedure is exemplified by extending the Szekeres class of colliding wave spacetimes with 2 additional real parameters. The obtained solution seems to feature a limiting case of a new type of impulsive waves, which are circularly polarised.
\end{abstract}



\section{Introduction}

Recent observations confirm the existence of gravitational waves (GW) emitted in strongly gravitating binaries, where the nonlinearity of the Einstein equations plays an important role. The observational data are convincingly reproducible with numerical models, however the performance of analytic descriptions of the strong gravity regime is still limited. In order to foster understanding and a creative utilisation of strong wave phenomena, a more analytical treatment is highly desirable.

A first step in this venture is surely the model of colliding plane GW, which is the simplest method to study nonlinear wave interactions analytically.
Therefore many features of nonlinearity as well as conceptual issues like focussing properties and arising singularities have been discussed on the basis of colliding plane waves so far.
A lot of exact solutions have been described along with solution generating techniques constructing solutions in the interaction region and deriving the shape of the incoming waves afterwards (cf. the overview of Griffiths \cite{Griffiths1991} or \cite{Griffiths_Podolsky2009}). Hauser and Ernst \cite{Hauser_ErnstI1989,Hauser_ErnstII1989,Hauser_ErnstIII1990,Hauser_ErnstIV1991,Hauser_Ernst_GerochConjecture2001} pioneered the search for a method to address the characteristic initial value problem and proved the existence and uniqueness of its solution.
Alekseev and Griffiths \cite{Alekseev_Griffiths2001,Alekseev_Griffiths2004} described a more practical procedure for both colliding gravitational and electromagnetic waves leading to integral equations for singular functions. Our treatment of the characteristic initial value problem features many similarities to this approach, but allows for additional transformations to integral equations for non-singular functions. It is expected to be better suited for approximations using spectral methods, but it is still too early to clearly compare the performance of the two approaches related to this goal. Finally we aim for new analytic solutions on the one hand and a systematic study of interaction properties of colliding plane GW depending on the initial data on the other hand. Concerning numerical evaluation, the inverse scattering method is complementary to the more common finite differencing schemes because the solution at a specified point can be calculated with high accuracy independent of its environment, especially without accumulating errors.

In this paper we consider purely gravitational plane waves with distinct wavefronts and arbitrary polarisation colliding in a Minkowski background. The corresponding spacetime features an orthogonally transitive two-dimensional group of isometries essentially reducing the Einstein equations to the hyperbolic Ernst equation. We make use of the strong formal analogy to axially symmetric and stationary spacetimes governed by the elliptic Ernst equation by formulating a `linear problem' (LP) in the Neugebauer form, cf. \cite{Neugebauer_Meinel2003}. Its solution can be represented by the solution of a Riemann-Hilbert problem (RHP) whose jump matrix is defined by the characteristic initial values. This procedure is known as the `inverse scattering method', cf. \cite{Belinski_Verdaguer2001} for a general introduction and \cite{Neugebauer_Meinel2003} for the axisymmetric analog of our particular case. Due to the inevitable non-analytic behaviour of the initial data on the wavefronts, the jump matrix is discontinuous. Adapting a general method of Vekua \cite{Vekua1967}, we implement a transformation to a continuous RHP (cRHP) which can be solved using integral equations for the non-singular additive jump of the cRHP solution. 

Because of the non-analytic behaviour of the initial data, the RHP does not uniquely define the solution to the LP and we need to impose regularity conditions of the LP coefficient matrices. The remaining degrees of freedom lead to families of solutions and therefore our procedure can also serve as solution generating technique. This is illustrated via the application of our method on the Szekeres class \cite{Szekeres1972} of colliding wave solutions resulting in a generalised class of exact spacetimes with 2 additional parameters. An interesting limiting case featuring `circularly polarised impulsive waves' seems to be included in this class.


\section{The characteristic initial value problem for colliding plane waves}

\subsection{Ernst equation}

We write the metric in the Szekeres-form \cite{Szekeres1972} with the parametrisation as given in \cite{Griffiths1991} :
\begin{equation}  
 \rmd s^2=2\rme^{-M}\rmd u\rmd v-\frac{2\rme^{-U}}{E+\bar{E}}\vert \rmd x+\rmi E\rmd y\vert^2. \label{metric}
\end{equation}
It contains the two real functions $M(u,v)$ and $U(u,v)$ and the complex Ernst potential $E(u,v)$ only depending on the two lightlike coordinates $u$ and $v$. The spacelike coordinates $x$ and $y$ parametrise the planes of symmetry.
The vacuum Einstein equations reduce to the essential relations 
\begin{eqnarray}
U_{uv}=U_{u}U_{v}, \label{Uuv}\\
(E+\bar{E})(2E_{uv}-U_u E_v-U_v E_u)=4 E_u E_v, \label{Euv} \\
(E+\bar{E})^2(2U_{uu}-U_u^2+2M_u U_u)=4 E_u \bar{E}_u , \label{Mu}\\
(E+\bar{E})^2(2U_{vv}-U_v^2+2M_v U_v)=4 E_v \bar{E}_v,  \label{Mv}
\end{eqnarray} 
where by coordinate indices $u$ and $v$ as well as $f$ and $g$ below we denote partial derivatives. Equation \eref{Uuv} has the general solution
\begin{equation} 
\rme^{-U}=f(u)+g(v)
\end{equation}
containing two arbitrary functions $f(u)$ and $g(v)$. In accordance to Griffiths \cite{Griffiths1991} we choose
\begin{eqnarray}
f=\mbox{$\frac12$} \;\;{\rm for}\; u\leq 0,\quad g=\mbox{$\frac12$} \;\;{\rm for}\; v\leq 0, \quad f'(0)=0=g'(0), \label{fgStart}
\end{eqnarray}
whereby \eref{Mu} and \eref{Mv} show that $f$ and $g$ are monotonically decreasing for $u,v\geq 0$. Using them as coordinates, \eref{Euv} becomes the hyperbolic Ernst equation
\begin{equation} \label{ErnstGl}
\Re(E)\left(2E_{fg}+\frac{E_f+E_g}{f+g}\right)=2 E_f E_g.
\end{equation}
Having determined $E$ and $U$ we can afterwards obtain the function $M$ by integration of the field equations \eref{Mu} and \eref{Mv}. Integrability is assured by the Ernst equation \eref{ErnstGl}. Together with $E$ also the function $E'=aE+\rmi b$ ($a,b\in\mathbb{R}$) is a solution to the Ernst equation \eref{ErnstGl}. We fix this freedom by demanding as connection to the Minkowski background the normalisation
\begin{equation}
E(\mbox{$\frac12$},\mbox{$\frac12$})=1. \label{NormErnst}
\end{equation}




\subsection{Spacetime regions}

\begin{figure}[ht]
\centering
\includegraphics[width=0.50\linewidth]{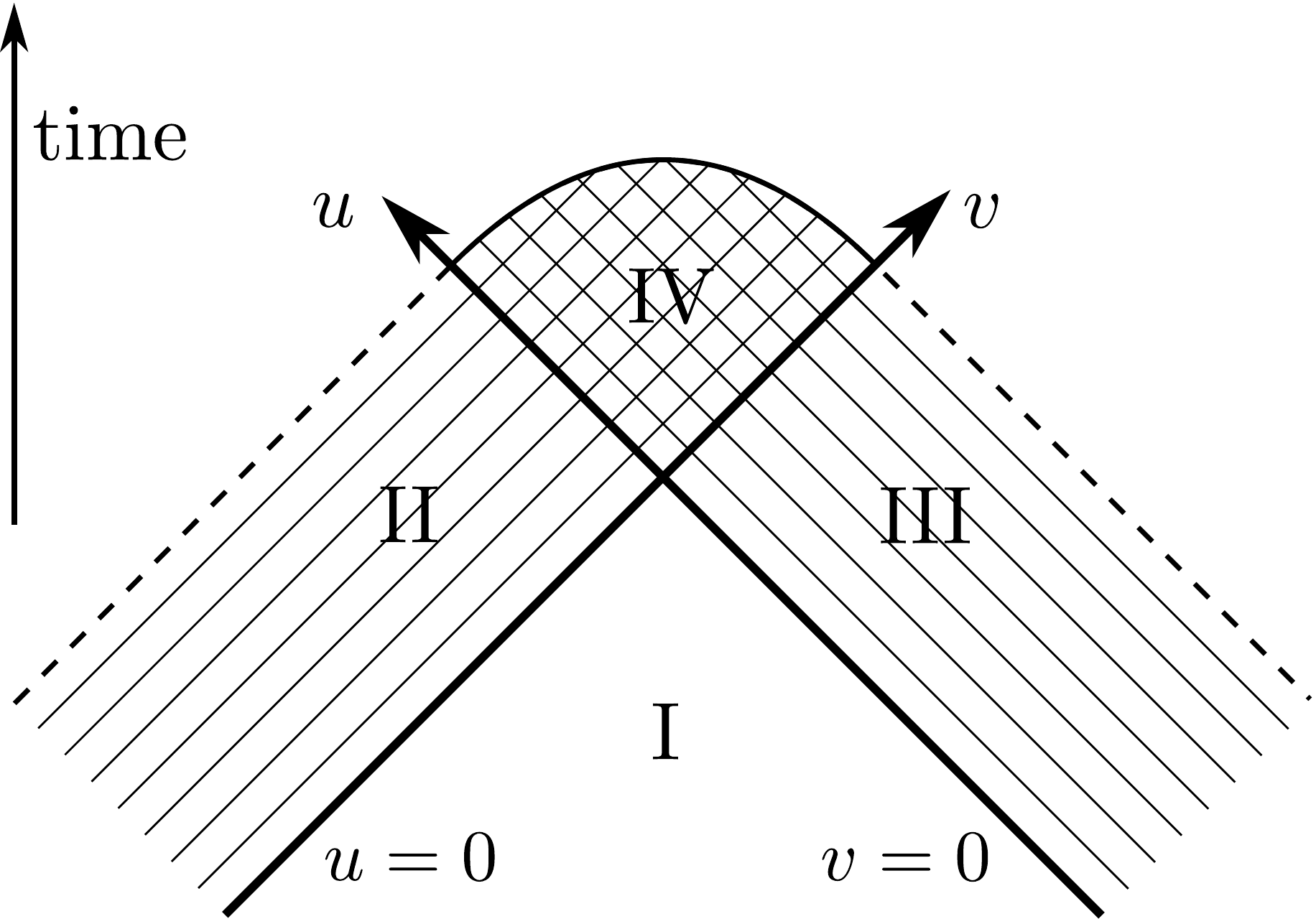}
\caption{Identification of the 4 spacetime regions of colliding GW adapted from Griffiths \cite{Griffiths1991}, see also \cite{Szekeres1972,Khan_Penrose1971}.}
\label{fig:regions}
\end{figure}

As illustrated in \fref{fig:regions}, it is appropriate to divide a colliding wave spacetime into four regions \cite{Khan_Penrose1971} with the following coordinate dependencies of the metric functions: 
\begin{eqnarray} \fl\begin{array}{rllll}
I: & u<0,v<0: \quad & E=1,      & M=0,           & \rme^{-U}=1, \\
II: & u\geq 0,v<0: & E(u,0), \;& M(u,0)=:M_{II}(u),  & \rme^{-U}=\frac12+f(u), \\
III: & u<0,v\geq 0: & E(0,v), & M(0,v)=:M_{III}(v),\;& \rme^{-U}=\frac12+g(v), \\
IV: & u\geq 0,v\geq 0: & E(u,v), & M(u,v),       & \rme^{-U}=f(u)+g(v). \\
\end{array}\end{eqnarray}

The physical interpretation is that on a Minkowski background (I) two plane waves propagate undisturbed in opposite direction (II and III, there is always a frame where the collision happens `head on') until their collision and nonlinear interaction(IV).

 Using the functions $f$ and $g$ as coordinates in the interaction region IV, the characteristic initial value problem of colliding plane GW corresponds to finding a solution $E$ to the Ernst equation \eref{ErnstGl} with given initial values $E(f,\frac12)$ and $E(\frac12,g)$ respecting the normalisation \eref{NormErnst}. Due to the fact that these boundaries are the characteristic curves of the Ernst equation, it is sufficient to provide  initial values of $E$ without additionally giving its derivatives.

As indicated by the $f+g$ in the denominator within \eref{ErnstGl} (and discussed in detail in \cite{Griffiths1991}), the colliding wave spacetime features a generic scalar curvature singularity at $f+g=0$ (solid curved line in \fref{fig:regions}). This can be understood by the mutual focussing properties of waves in GR. For a large variety of exceptional cases this singularity is replaced by a Killing-Cauchy horizon, but this horizon is conjectured to be unstable \cite{Yurtsever1989}. For collinearly polarised waves this instability has been rigorously proven \cite{Griffiths2005}. The regions II and III are confined by coordinate degeneracies on lightlike hypersurfaces (dashed lines in \fref{fig:regions}). They can be identified with the points $-f=g=\frac12$ and $f=-g=\frac12$ and inherit their singular character. Nevertheless, for the vacuum case considered here they are no scalar curvature singularities on their own and so the term `fold singularity' has been established to indicate their topological character. 

\subsection{Colliding wave conditions}

Using the junction conditions of O'Brien and Synge \cite{O’Brien_Synge1952} for lightlike boundaries (shown to be appropriate by Robson \cite{Robson_1973}) the metric has to meet the following demands:
\begin{eqnarray} \label{ÜB}
E,M \in C^0,\quad U \in C^1.
\end{eqnarray} 
This allows us to perform $C^1$-transformations $u\to u'(u)$ and $v\to v'(v)$ to arrange 
\begin{eqnarray} \label{fgWahl}
f=\mbox{$\frac12$}-(c_1 u)^{n_1}\Theta(u), \quad\quad g=\mbox{$\frac12$}-(c_2 v)^{n_1}\Theta(v),
\end{eqnarray}
where $\Theta(\cdot)$ is the Heaviside step function and $c_{1/2}$ can be interpreted as magnitudes of the waves. Alternatively, we may use such $C^1$-transformations to achieve $M_{II}(u)=0$  and $M_{III}(v)=0$. Then $f(u)$ and $g(v)$ are determined by the field equations \eref{Mu}, \eref{Mv} and the junction conditions \eref{fgStart} with the Minkowski background. Also in this case the exponents $n_{1/2}$ describe the first order behaviour of $f(u)$ and $g(v)$ because they cannot be changed by $C^1$-transformations. The field equations \eref{Mu}, \eref{Mv} impose the restriction
\begin{eqnarray} 
n_{1/2} \geq 2, \label{nRange}
\end{eqnarray} 
where $n_{1/2} = 2$ implies an impulsive wavefront. Furthermore, together with the continuity of $M$, they lead to the so-called `colliding wave conditions' for $E$ first formulated by Hauser and Ernst \cite{Hauser_ErnstI1989}, which within the normalisation \eref{NormErnst} have the form
\begin{eqnarray}
\lim_{\left(f,g\right)\rightarrow(\frac12,\frac12)} \left[(\mbox{$\frac12$}-f)E_f\bar{E}_f\right]=2k_1, \quad
\lim_{\left(f,g\right)\rightarrow(\frac12,\frac12)} \left[(\mbox{$\frac12$}-g)E_g\bar{E}_g\right]=2k_2, \label{cwc}
\end{eqnarray}
with 
\begin{eqnarray} \label{kWerte}
k_{1/2}=1-\frac{1}{n_{1/2}},  \quad\quad \frac12\leq k_{1/2}<1.
\end{eqnarray}
In the context of the characteristic IVP the colliding wave conditions are a matter of choosing suitable initial values for $E$ featuring divergent derivatives at $(f=\frac12,g=\frac12)$.

\section{Inverse scattering method for collinear polarisation}

\subsection{Scheme of inverse scattering}

In the course of the inverse scattering method, the nonlinear Ernst equation is expressed as the integrability condition of a system of linear partial differential equations, the so-called linear problem (LP). This system is in turn solved by constructing an appropriate Riemann-Hilbert problem (RHP) which has the same solution. This solution's inner and outer limits at a contour in the complex plane of a spectral parameter are through the RHP related by a purely multiplicative matrix-valued jump. In correspondence with the boundaries of the IVP, the contour of the RHP has to be chosen as two specific parts of the real axis (cf. \fref{Kontur}). It is possible to construct a single closed contour by continuation of these parts, whereby we have to set the jump matrix to $\mathbb{1}$ on the added parts. Regrettably, the singularity in the derivatives of the Ernst potential demanded by the colliding wave conditions \eref{cwc} leads to a jump matrix which tends to a finite value different from $\mathbb{1}$ at the ends of the  initial contour. Therefore, the RHP is discontinuous and its solutions feature singularities at the ends of the initial contour. In fact, there are even two different appearances of these singularities, so that an ambiguity in the RHP solution arises at each of the two initial contours and we end up with 4 different solutions of the RHP.
 In this article we perform a transformation to a continuous Riemann-Hilbert problem (cRHP) which clarifies these ambiguities and proves the existence of the RHP solutions. The cRHP is also supposed to be better suited for a numerical treatment  than the initial (discontinuous) RHP.

\begin{figure}[ht]
\centering
\includegraphics[width=\linewidth]{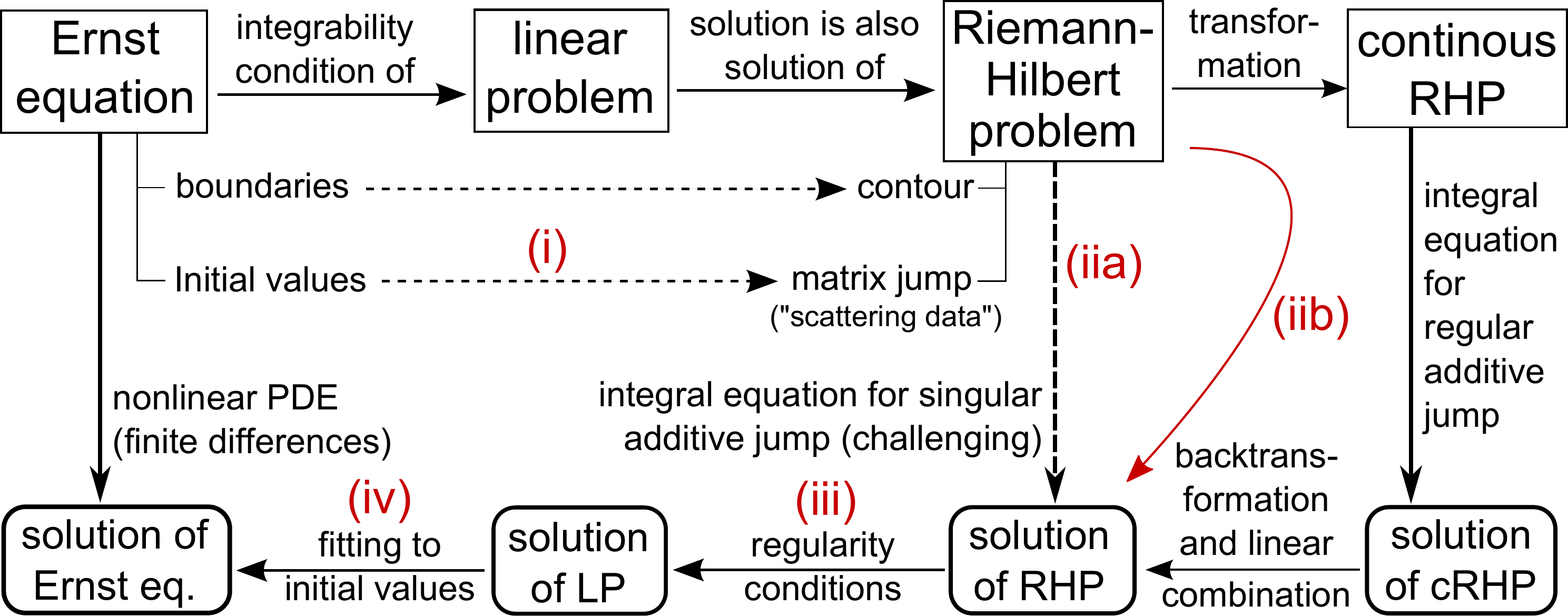}
\caption{Scheme of the inverse scattering method with additional transformation to a continuous Riemann-Hilbert problem}
\label{Lösungsschema}
\end{figure} 

In the course of our procedure, for given initial data $E(f,\frac12)$ and $E(\frac12,g)$ the following four steps indicated in \fref{Lösungsschema} have to be carried out: 
\begin{enumerate}
\item Translating the initial data into the jump matrix by solving a system of ordinary differential equations (ODE). For special cases an analytical treatment is possible.
\item Solving the Riemann-Hilbert problem, in the general case by expansion of an additive jump function in Chebyshev polynomials
\begin{enumerate}
\item via integral equations for singular additive jumps with bad numerical properties
\item via transformation to the cRHP and its integral equations for regular additive jumps with better numerical properties
\end{enumerate}
\item Evaluating regularity conditions, which assure that the RHP solution fulfills the LP. These are purely algebraic equations to determine the linear combination coefficients of the RHP's 4 basic solutions.
\item Fixing the remaining degrees of freedom to adapt the solution to its initial data
\end{enumerate}
The whole process of the inverse scattering method shall be illustrated by examining the case of collinearly polarised gravitational waves. The contour of the collinearly polarised case will be directly transferred to the RHP for GW with arbitrary polarisation.

\subsection{Linear problem for collinear polarisation}

Within the Newman-Penrose formalism the singular waves in the spacetime regions II and III are described by the complex Weyl tensor components $\Psi_0(v)$ and $\Psi_4(u)$ respectively. For linearly polarized initial waves the phases of these components are constant in region II and III. If these constant phases are even identical, the metric can be diagonalised containing only a real Ernst potential. This very special setup is called the collision of collinearly polarised GW.

The general solution for collinear polarisation has been derived by Hauser and Ernst \cite{Hauser_ErnstI1989} in terms of generalized Abel transformations. It has been reformulated in order to obtain an initial point for the generalisation to arbitrary polarisation \cite{Hauser_ErnstII1989}. We will follow the same line here to establish our methods through the collinear case.



The LP for collinear polarisation is to find the function $\Phi^{\|\rm LP}(f,g;\lambda)$ satisfying
\begin{eqnarray} \label{LP}
\begin{array}{c}
 \Phi_f^{\|\rm LP} =(1+\lambda)A\Phi^{\|\rm LP}\\ 
 \Phi_g^{\|\rm LP} =(1+\frac{1}{\lambda})B\Phi^{\|\rm LP}
\end{array},\quad\quad
\lambda=\sqrt{\frac{k-g}{k+f}},
\end{eqnarray}
where $A$ and $B$ are real functions of $f$ and $g$ whereas $k$ is an independent spectral parameter, which enters the equations through the spectral parameter $\lambda$ depending on $f$, $g$ and $k$. The partial derivatives $\partial_f$ and $\partial_g$ are taken with constant $k$ rather than constant $\lambda$. The solution $\Phi^{\|\rm LP}$ can be thought of either as a function on the extended complex $\lambda$-plane $\mathbb{C}_\lambda:=\mathbb{C}\cup\{\infty\}$ or as a function of $k$ defined on a two-sheeted Riemann surface with branch cut along the segment $[-f,g]$, a twofold covering of the (in the same sense extended) complex $k$-plane, cf. \fref{Konturk}. We call the sheet with $\lambda\to 1$ for $k\to\infty$ the upper one and the sheet with $\lambda\to -1$ for $k\to\infty$ the lower one.

The integrability conditions $\Phi_{fg}^{\|\rm LP}=\Phi_{gf}^{\|\rm LP}$ of the LP \eref{LP} assure the existence of a potential $\psi^\|$ fulfilling
\begin{eqnarray}
\psi_f^\|=A, \\
\psi_g^\|=B, \\
2\psi_{fg}^\|+\frac{\psi_f^\|+\psi_g^\|}{f+g}=0. \label{linErnst}
\end{eqnarray}
The linear equation \eref{linErnst} is indeed the Euler-Poisson-Darboux equation, which one can derive from the Ernst equation in case of real $E$ by setting 
\begin{equation}
\psi^\|=\mbox{$\frac12$}\ln E. \label{DefPsippE}
\end{equation}

The LP solution $\Phi^{\|\rm LP}$ is only defined up to multiplication with a function of $k$. We now fix this freedom by demanding the normalisation
\begin{equation}
\Phi^{\|\rm LP}\left(\mbox{$\frac12$},\mbox{$\frac12$}\right)=1 \quad\quad \forall k. \label{NormPhiLP}
\end{equation}
Here and in the rest of the paper we use the following convention for an arbitrary function $F$ depending on $f$, $g$ and $\lambda$: Where $F$ is displayed with 2 arguments as in \eref{NormPhiLP}, these should be understood as the values of $f$ and $g$, but where $F$ is displayed with a single argument as in \eref{PhipLPminEins}, this should be taken as the value of $\lambda$. Evaluating the LP \eref{LP} at $\lambda=-1$ and $\lambda=1$ equation \eref{NormPhiLP} easily leads to 
\begin{eqnarray} 
  \Phi^{\|\rm LP}(-1)=1,     \label{PhipLPminEins}\\ 
  \Phi^{\|\rm LP}(1)=e^{-2\psi^\|(\frac12,\frac12)}e^{2\psi^\|}.
\end{eqnarray}
The Ernst potential with the normalisation \eref{NormErnst} is therefore given by
\begin{eqnarray} 
  E=\Phi^{\|\rm LP}(1). \label{DefPsippPhi}
\end{eqnarray}

\subsection{Riemann-Hilbert problem for collinear polarisation}

The RHP connected to \eref{LP} is to find a function ${\Phi}^\|(f,g;\lambda)$ which is analytic everywhere in the complex Riemann $k$-surface except on the contour $\Gamma^{(k)}$, where it has a jump described by the equation
\begin{eqnarray} \label{Sprung}
{\Phi}_+^\|=\alpha(k){\Phi}_-^\|.
\end{eqnarray}
Herein ${\Phi}_+^\|$ is the inner (left to the contour) and ${\Phi}_-^\|$ the outer (right to the contour) limit of ${\Phi}^\|$. In addition, we fix the freedom of multiplying ${\Phi}^\|$ with a function of $f$ and $g$ by demanding the normalisation
\begin{eqnarray} \label{NormPhiRHP}
  {\Phi}^\|(-1)=1 \quad\quad \forall f,g.    
\end{eqnarray}

The derivatives of ${\Phi}^\|$ with respect to the coordinates can be written as
\begin{eqnarray} 
 {\Phi}^\|_f=({\Phi}^\|_f)_{g,\lambda\,=\,{\rm const}}+{\Phi}^\|_{\lambda}\lambda_f, \quad 
 {\Phi}^\|_g=({\Phi}^\|_g)_{f,\lambda\,=\,{\rm const}}+{\Phi}^\|_{\lambda}\lambda_g
\end{eqnarray}
with the partial derivatives
\begin{eqnarray} 
 \lambda_f=\frac{\lambda}{2(f+g)}(\lambda^2-1), \quad \lambda_g=\frac{1}{2(f+g)\lambda}(\lambda^2-1),
\end{eqnarray}
which are singular at $\lambda=\infty$ and $\lambda=0$ respectively. Therefore a power series   expansion of ${\Phi}^\|$ at $\lambda=\infty$ or rather $\lambda=0$ leads to 
\begin{eqnarray} 
\begin{array}{c} 
{\Phi}^\|_f({\Phi}^\|)^{-1}=c_{(-1)}(f,g)\lambda+c_{(0)}(f,g)+c_{(1)}(f,g)\lambda^{-1}+\dots, \\
{\Phi}^\|_g({\Phi}^\|)^{-1}=d_{(-1)}(f,g)\lambda^{-1}+d_{(0)}(f,g)+d_{(1)}(f,g)\lambda+\dots
\end{array}
\end{eqnarray}
Because the multiplicative jump $\alpha$ is a function only depending on the spectral parameter $k$, we can deduce by calculating derivatives of \eref{Sprung} that the terms ${\Phi}^\|_f({\Phi}^\|)^{-1}$ and ${\Phi}^\|_g({\Phi}^\|)^{-1}$ exhibit no jump on the contour $\Gamma^{(k)}$. Within the treatment of the cRHP we even show that there exists a solution of the RHP with ${\Phi}^\|_f({\Phi}^\|)^{-1}$ holomorphic in $\mathbb{C}_{\lambda}\backslash\{\infty\}$ and ${\Phi}^\|_g({\Phi}^\|)^{-1}$ holomorphic in $\mathbb{C}_{\lambda}\backslash\{0\}$. In this case, Liouville's theorem demands $c_{(i)}(f,g)=0=d_{(i)}(f,g) \,\forall i\geq 1$ and with the normalisation \eref{NormPhiRHP} we get the LP \eref{LP}.

\begin{figure}[ht!]
\centering
\includegraphics[width=0.85\linewidth]{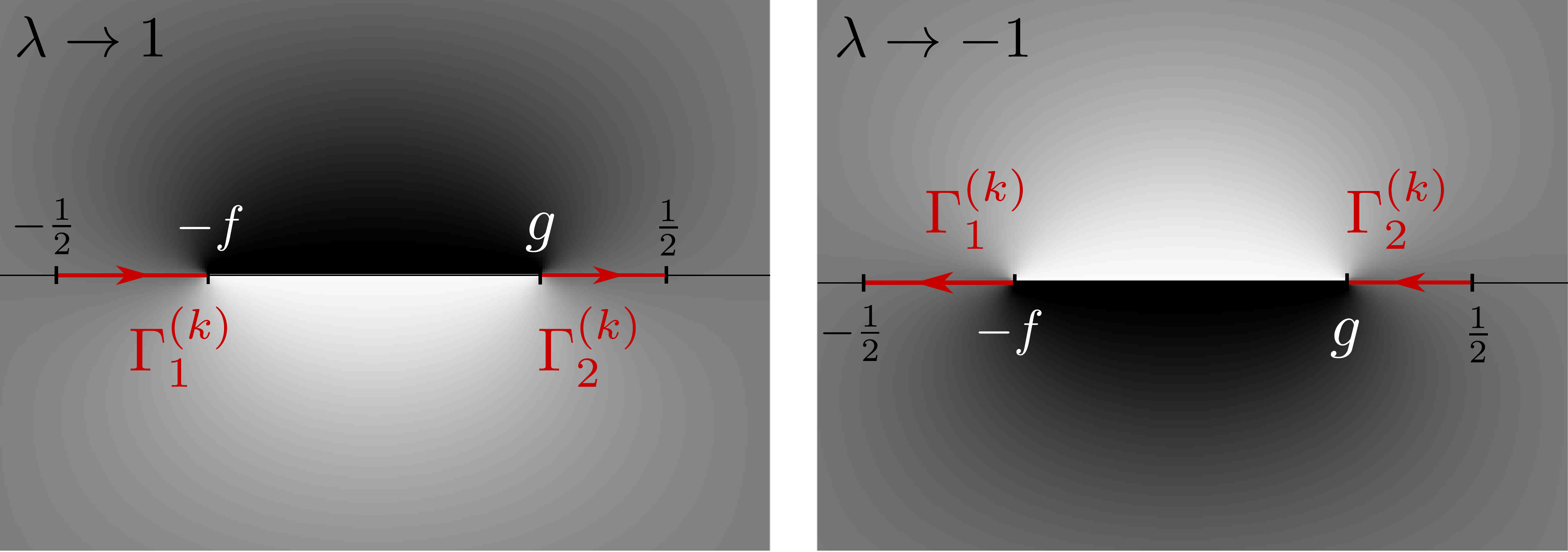}
\caption{The 2 parts $\Gamma^{(k)}_1$ and $\Gamma^{(k)}_2$ of the contour $\Gamma^{(k)}$ in the upper (left) and lower (right) sheet of the two-sheeted Riemann $k$-surface. At the branch cut $[-f,g]$ bright area is connected to bright area and dark area to dark area.}
\label{Konturk}
\end{figure}

As illustrated in \fref{Konturk} we choose the contour $\Gamma^{(k)}$ in the $k$-surface to consist of a first part $\Gamma^{(k)}_1$ directed from $k=-\frac12$ in the upper sheet through the branch point $k=-f$ to $k=-\frac12$ in the lower sheet and a second part $\Gamma^{(k)}_2$ directed from $k=\frac12$ in the lower sheet through the branch point $k=g$ to $k=\frac12$ in the upper sheet.

\begin{figure}[ht!]
\centering
\includegraphics[width=0.65\linewidth]{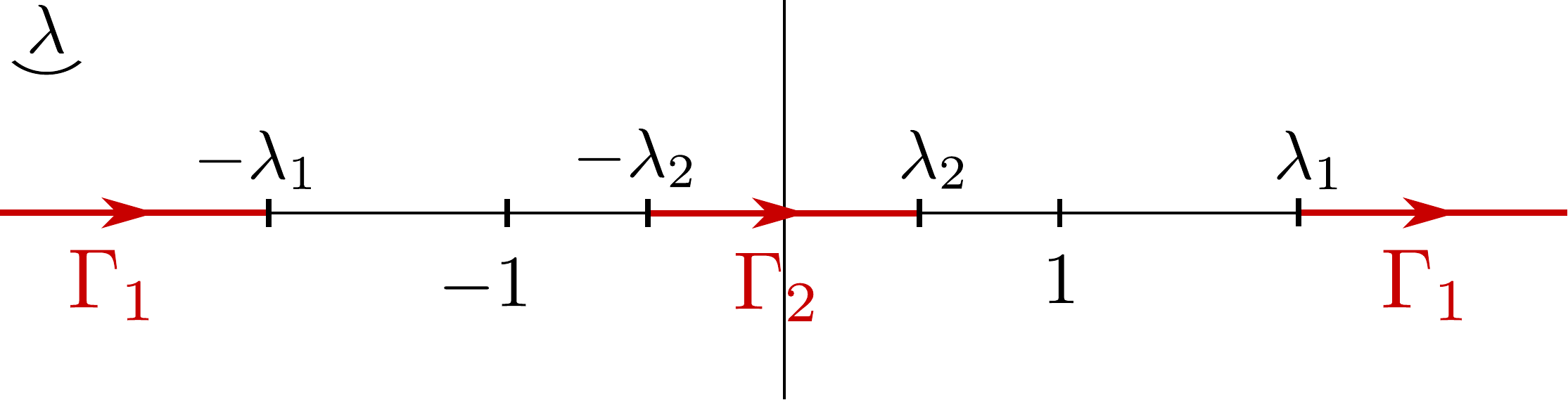}
\caption{The 2 parts $\Gamma_1$ and $\Gamma_2$ of the contour $\Gamma$ in the $\lambda$-plane.}
\label{Kontur}
\end{figure}

By setting $k=\pm \frac12$ we get the contour endpoints in the $\lambda$-plane (cf. \fref{Kontur}):
\begin{equation}
\lambda_1=\sqrt{\frac{\frac12+g}{\frac12-f}}, \quad\quad 
\lambda_2=\sqrt{\frac{\frac12-g}{\frac12+f}}.
\end{equation}
They lie on the real axis and satisfy $\lambda_1>1>\lambda_2>0$. The contour $\Gamma$ in the $\lambda$-plane is divided into $\Gamma_1$ corresponding to $\Gamma^{(k)}_1$ and $\Gamma_2$ corresponding to $\Gamma^{(k)}_2$. The first part $\Gamma_1$ is directed from $\lambda_1$ through $\lambda=\infty$ to $-\lambda_1$ and the second part $\Gamma_2$ is directed from $-\lambda_2$ through $\lambda=0$ to $\lambda_2$.

The contour vanishes for $f=\frac12=g$, which leads with \eref{NormPhiRHP} to a reproduction of the LP normalisation ${\Phi}^\|(\frac12,\frac12)=1$. Therefore normalised RHP solutions with ${\Phi}^\|_f({\Phi}^\|)^{-1}$ holomorphic in $\mathbb{C}_{\lambda}\backslash\{\infty\}$ and ${\Phi}^\|_g({\Phi}^\|)^{-1}$ holomorphic in $\mathbb{C}_{\lambda}\backslash\{0\}$ are also solutions ${\Phi}^{\|\rm LP}$ of the normalised LP. 

In the collinearly polarised case we can rewrite \eref{Sprung} as the additive jump equation
\begin{eqnarray} \label{SprungAdd}
\ln{\Phi}^\|_+-\ln{\Phi}^\|_-=\ln\alpha=:\rmi\mu^\|.
\end{eqnarray}
The solution can be given in terms of a Cauchy integral in the $\lambda$-plane as
\begin{eqnarray} \label{lnPhi}
\ln{\Phi}^\|=\frac{1}{2\pi}\int_{\Gamma}\left(\frac{1}{\lambda'-\lambda}-\frac{1}{\lambda'+1}\right)\mu^\|(k')\rmd\lambda',
\end{eqnarray}
where the second term under the integral assures  the normalisation \eref{NormPhiRHP}.
Evaluating \eref{lnPhi} at $\lambda=1$ leads with \eref{DefPsippE} and \eref{DefPsippPhi} to 
\begin{eqnarray} \label{psiLineInt}
\psi^\|=\frac{1}{2\pi}\int_{-\frac12}^{-f}\frac{\mu^\|_1(k')\rmd k'}{\sqrt{(k'-g)(k'+f)}}
     -\frac{1}{2\pi}\int_{g}^{\frac12}\frac{\mu^\|_2(k')\rmd k'}{\sqrt{(k'-g)(k'+f)}},
\end{eqnarray}
where $\mu^\|_{1/2}=\mu^\||_{\Gamma_{1/2}^{(k)}}$. With the index `$1/2$' we denote a statement holding both for index $1$ and for index $2$ inserted throughout the entire expression. For $f=\frac12$ the second line of the linear problem \eref{LP} reads
\begin{eqnarray} \label{LPfeinhalb}
\left(\ln\Phi^\|\left(\mbox{$\frac12$},g\right)\right)_g=(1+\lambda^{-1})\psi^\|_g\left(\mbox{$\frac12$},g\right).
\end{eqnarray}
With \eref{NormPhiLP} the integration from $g'=\frac12$ to $g'=g$ yields
\begin{eqnarray} \label{fgleicheinhalb}
\ln\Phi^\|(\mbox{$\frac12$},g)
   =\psi^\|(\mbox{$\frac12$},g)-\int_g^{\frac12}\sqrt{\frac{k+\frac12}{k-g'}}\psi^\|_{g'}\rmd g'.
\end{eqnarray}
From this the additive jump on $\Gamma^{(k)}_2$ can be shown to be
\begin{eqnarray} \label{mu2}
\mu^\|_2=-i(\ln\Phi^\|_+-\ln\Phi^\|_-)=2\sqrt{\mbox{$\frac12$}+k}\int_{k}^{\frac12}\rmd g'\frac{\psi^\|_{g'}(\frac12,g')}{\sqrt{g'-k}}
\end{eqnarray}
and analogously the additive jump on $\Gamma^{(k)}_1$ is
\begin{eqnarray} \label{mu1}
\mu^\|_1=-i(\ln\Phi^\|_+-\ln\Phi^\|_-)=-2\sqrt{\mbox{$\frac12$}-k}\int_{-k}^{\frac12}\rmd f'\frac{\psi^\|_{f'}(f',\frac12)}{\sqrt{k+f'}}.
\end{eqnarray}
Note that these jump functions both are defined on the interval $[-\frac12,\frac12]$, but for given $f$ and $g$ only the values of $\mu^\|_{1}$ on $\Gamma^{(k)}_1$ and the values of $\mu^\|_{2}$ on $\Gamma^{(k)}_2$ appear in \eref{psiLineInt}. Obviously real initial values $\psi^\|_{f}(f,\frac12)$, $\psi^\|_{g}(\frac12,g)$ lead to real $\mu^\|_{1/2}$ and via \eref{psiLineInt} to a real solution $\psi^\|$. The combination of \eref{mu2}, \eref{mu1} and \eref{psiLineInt} constitutes the general solution of the IVP for collinearly polarised colliding plane waves. 

In \cite{Hauser_ErnstI1989} this solution is derived with a generalised Abel transformation. Furthermore, a RHP similar to \eref{Sprung} is presented, where the spectral parameter lies in a simple complex plane. Its solution $\Phi_H$ is related to $\Phi^\|$ by
\begin{eqnarray} \label{Beziehung_2}
\ln\Phi^\|(k)=-(k+f)\lambda\Phi_H(2k)+\psi^\|
\end{eqnarray}
and it uses the jump functions
\begin{eqnarray} 
g_3(\sigma)=\frac{\mu^\|_1(\sigma/2)}{2\sqrt{1-\sigma}},\quad\quad g_2(\sigma)=\frac{\mu^\|_2(\sigma/2)}{2\sqrt{1+\sigma}}.
\end{eqnarray}

\section{Inverse scattering method for arbitrary polarisation}

\subsection{Linear problem for arbitrary polarisation}

The LP for colliding plane waves of arbitrary polarisation is to find the matrix $\Phi^{\rm LP}(f,g;\lambda)$ satisfying
\begin{equation} \label{algLP}
\begin{array}{c}
 \Phi^{\rm LP}_f =U\Phi^{\rm LP}\\ 
 \Phi^{\rm LP}_g =V\Phi^{\rm LP}
\end{array},\quad
U=
\left(\begin{array}{cc}
A & \lambda  A \\ 
\lambda \bar{A}  & \bar{A}             \\ 
\end{array}\right),\quad
V=
\left(\begin{array}{cc}
B & \lambda^{-1} B  \\ 
\lambda^{-1} \bar{B}  & \bar{B}             \\ 
\end{array}\right).
\end{equation}
Herein $A$ and $B$ are complex functions of $f$ and $g$ and the spectral parameter $\lambda$ is defined as in \eref{LP}. In addition, we fix the freedom of right-multiplying a matrix function of $k$ to the solution, $\Phi^{\rm LP}\rightarrow\Phi^{\rm LP} C(k)$, by demanding the normalisation
\begin{eqnarray} \label{LPNorm}
\Phi^{\rm LP}(\mbox{$\frac12$},\mbox{$\frac12$})=\left(\begin{array}{cc} 1 & -1 \\ 1 & 1 \end{array}\right) \quad \forall k.
\end{eqnarray}

The integrability conditions of the LP \eref{algLP} assure the existence of a potential $E$ fulfilling the equations
\begin{eqnarray}
\frac{E_f}{E+\bar{E}}=A, \quad\quad
\frac{E_g}{E+\bar{E}}=B \label{ABdurchE}
\end{eqnarray}
and the Ernst equation \eref{ErnstGl}.

To motivate the design of the RHP, we derive some properties of the LP. We proceed analogous to \cite{Neugebauer_Meinel2003}. 
Using the Pauli matrices
\begin{eqnarray}
\sigma_1=\left(\begin{array}{cc} 0 & 1 \\ 1 & 0 \end{array}\right), \quad
\sigma_3=\left(\begin{array}{cc} 1 & 0 \\ 0 & -1 \end{array}\right)
\end{eqnarray}
we can state the following relations between the matrices  $U(\lambda)$ and $V(\lambda)$ and their values at $-\lambda$ and $\bar{\lambda}$ respectively:
\begin{eqnarray} 
\sigma_3 W(-\lambda) \sigma_3=W(\lambda)=\sigma_1 \bar{W}(\bar{\lambda}) \sigma_1
\quad {\rm with}\quad &W=U,V. 
\end{eqnarray}
Therefore, from a given column vector $v(\lambda)$ solving the LP we can derive the new solutions $\sigma_3 v(-\lambda)$ and $\sigma_1 \bar{v}(\bar{\lambda})$.  Hence we can construct a matrix solution of the LP, 
\begin{eqnarray} 
\fl \Phi^{\rm LP}=\left(\begin{array}{cc} \varphi^{\rm LP}(\lambda) & -\varphi^{\rm LP}(-\lambda) \\ \bar{\varphi}^{\rm LP}(\bar{\lambda}) & \bar{\varphi}^{\rm LP}(-\bar{\lambda}) \end{array}\right):=\left(v(\lambda)+\sigma_1 \bar{v}(\bar{\lambda}),-\sigma_3 [v(-\lambda)+\sigma_1 \bar{v}(-\bar{\lambda})]\right),
\label{Matrixlösung}
\end{eqnarray}
depending only on a single scalar function $\varphi^{\rm LP}$ which we will call the scalar solution of the LP. The representation \eref{Matrixlösung} is consistent with the normalisation \eref{LPNorm} providing that 
\begin{equation}
\varphi^{\rm LP}(\mbox{$\frac12$},\mbox{$\frac12$})=1\quad\forall k, \label{NormChi}
\end{equation}
and so we will assume the matrix solution of the LP and the RHP later on to have this structure \eref{Matrixlösung}, which we will abbreviate by saying `$\Phi^{\rm LP}$ is in normal form with the scalar function $\varphi^{\rm LP}$'. 

The $(1,1)$-elements of the LP equations \eref{algLP} yield for $\lambda=1$ with \eref{ABdurchE}:
\begin{eqnarray}
\fl\varphi^{\rm LP}_f(1)=\frac{E_f}{E+\bar{E}}(\varphi^{\rm LP}(1)+\bar{\varphi}^{\rm LP}(1)), \quad\quad 
\varphi^{\rm LP}_g(1)=\frac{E_g}{E+\bar{E}}(\varphi^{\rm LP}(1)+\bar{\varphi}^{\rm LP}(1)) \label{ChiF}.
\end{eqnarray}
Integration of the absolute values of \eref{ChiF} leads to $\varphi^{\rm LP}(1)=aE+\rmi b$,
 $a,b \in\mathbb{R}$. Considering \eref{NormChi} the Ernst potential with the normalisation \eref{NormErnst} is given by $E:=\varphi^{\rm LP}(1)$. Evaluation of the LP at $\lambda=-1$ in the same way leads to $\varphi^{\rm LP}(-1)=1$.
 
With the identity $(\ln \det M)_x=\Tr(M_x M^{-1})$, holding for an arbitrary square matrix $M$ as well as the normalisations \eref{NormErnst} and \eref{NormChi} of the Ernst potential and $\varphi^{\rm LP}$ we can derive from the LP the relation
\begin{equation}
\det\Phi^{\rm LP}=\bar{\varphi}^{\rm LP}(\bar{\lambda})\varphi^{\rm LP}(-\lambda)+\bar{\varphi}^{\rm LP}(-\bar{\lambda})\varphi^{\rm LP}(\lambda)=E+\bar{E} \qquad \forall f,g,k. \label{DetPhi}
\end{equation}
In particular it states that the determinant of the LP solution is a function depending only on the coordinates $f$ and $g$.

\subsection{Riemann-Hilbert problem for arbitrary polarisation}

The RHP for arbitrary polarisation is to find the matrix $\Phi(f,g;\lambda)$ analytic in $\mathbb{C}_{\lambda}\backslash\Gamma$ and satisfying on $\Gamma$ the jump equation
\begin{eqnarray}
\Phi_+= \Phi_- J(k). \label{RHPSprung} 
\end{eqnarray}
The jump matrix $J(k)$ has the form
\begin{eqnarray}
J(k)=\left(\begin{array}{cc} \alpha(k) & \beta(k) \\  -\beta(k) & \bar{\alpha}(k) \end{array}\right) \quad\rm{with}\quad \beta \in \mathbb{R},\; \bar{\alpha}\alpha+\beta^2=1,     \label{CForm} 
\end{eqnarray}
exhibiting only one complex degree of freedom $\alpha$. It is sufficient to consider the jump matrix to be identic in both sheets of the $k$-surface and so we set $J(-\lambda)=J(\lambda)$. Fixing the freedom of left-multiplying an arbitrary matrix $M(f,g)$ we demand the normalisation
\begin{equation}
\Phi(-1)=\left(\begin{array}{cc} 1 & -E \\ 1  & \bar{E} \end{array}\right). \label{NormRHP}
\end{equation}

As the investigation of the cRHP will show, there exists a solution $\Phi$ of the RHP in normal form \eref{Matrixlösung} with a scalar function $\varphi$ and fulfilling the generic regularity conditions 
\begin{eqnarray}\begin{array}{l}
{\Phi}_f\Phi^{-1} \mbox{ holomorphic in } \mathbb{C}_{\lambda}\backslash\{\infty\}, \\ {\Phi}_g\Phi^{-1} \mbox{ holomorphic in } \mathbb{C}_{\lambda}\backslash\{0\}.\end{array}\label{holoEigensch}
\end{eqnarray}
Similar to the collinearly polarised case, a power series expansion in $\lambda$ leads to 
\begin{eqnarray} 
\fl {\Phi}_f({\Phi})^{-1}=M_{(-1)}(f,g)\lambda+M_{(0)}(f,g), \quad\quad 
{\Phi}_g{{\Phi}}^{-1}=N_{(-1)}(f,g)\lambda^{-1}+N_{(0)}(f,g). \label{UVstruktur}
\end{eqnarray}

The representation \eref{Matrixlösung} is consistent with the normalisation \eref{NormRHP} if
\begin{eqnarray}
\varphi(1)=E, \label{DefE}\\
\varphi(-1)=1 \quad\forall f,g. \label{NormChiRHP}
\end{eqnarray}
Evaluating \eref{UVstruktur} at $\lambda=-1$ and $\lambda=1$ using \eref{DefE} and \eref{NormChiRHP} leads exactly to the LP \eref{algLP}. 

For $f=\frac12=g$ the contour $\Gamma$ vanishes and so the solution $\Phi$ is independent of $k$. Considering \eref{NormChiRHP} we find the normalisation \eref{LPNorm} of the LP reproduced. In consequence, normalised RHP solutions with ${\Phi}_f\Phi^{-1}$ holomorphic in $\mathbb{C}_{\lambda}\backslash\{\infty\}$ and ${\Phi}_g\Phi^{-1}$ holomorphic in $\mathbb{C}_{\lambda}\backslash\{0\}$
are also solutions $\Phi^{\rm LP}$ of the normalised LP. Equation \eref{NormChiRHP} can be regarded as the single normalisation condition for the RHP solution in normal form with $\varphi$, whereas \eref{DefE} defines the associated solution of the Ernst equation with the required normalisation \eref{NormErnst}. In order to match the initial values with this Ernst potential, the jump functions $\alpha$ and $\beta$ have to be determined from these initial values $E(f,g=\frac12)$ and $E(f=\frac12,g)$.

For $\beta=0$ the RHP \eref{RHPSprung} reduces to the collinearly polarised case   with $\mu^\|:=-\rmi\ln\alpha$, $\Phi^\|:=\varphi=\bar{\varphi}(\bar{\lambda})$ and $\psi^\|:=\frac12\ln E$.

\subsection{Calculation of the jump matrix from initial data}

At first we want to note that reading off the values of the scalar solution $\varphi$ at $-\lambda$ or $\bar{\lambda}$ implies in our setup changing the side of the contour:
\begin{eqnarray}
\left[\varphi(-\lambda)\right]_{+/-}=\varphi_{-/+}(-\lambda), \quad
\left[\varphi(\bar{\lambda})\right]_{+/-}=\varphi_{-/+}(\bar{\lambda})=\varphi_{-/+}(\lambda).
\end{eqnarray}

Remembering $J(-\lambda)=J(\lambda)$ we can convert the jump equation \eref{RHPSprung} to
\begin{eqnarray}
\fl\alpha = \frac{\bar{\varphi}_-(\lambda)\varphi_+(-\lambda)+\varphi_+(\lambda)\bar{\varphi}_-(-\lambda)}{\bar{\varphi}_+(\lambda)\varphi_+(-\lambda)+\bar{\varphi}_-(-\lambda)\varphi_-(\lambda)}, \quad
\beta = \frac{\bar{\varphi}_+(\lambda)\varphi_+(\lambda)-\varphi_-(\lambda)\bar{\varphi}_-(\lambda)}{\bar{\varphi}_+(\lambda)\varphi_+(-\lambda)+\bar{\varphi}_-(-\lambda)\varphi_-(\lambda)}. \label{alphabetachi}
\end{eqnarray}
It is now convenient to introduce $\chi=\bar{\varphi}(\bar{\lambda})$. Using $\alpha_{1/2}:=\alpha|_{\Gamma_{1/2}^{(k)}}$ and $\beta_{1/2}:=\beta|_{\Gamma_{1/2}^{(k)}}$ the evaluation of \eref{alphabetachi} at $\lambda=\infty$ and $\lambda=0$ yields  
\begin{eqnarray}
\alpha_1 = \frac{2\chi_+(\infty)\varphi_+(\infty)}{|\varphi_+(\infty)|^2+|\chi_+(\infty)|^2}, \quad &
\beta_1 = \frac{|\varphi_+(\infty)|^2-|\chi_+(\infty)|^2}{|\varphi_+(\infty)|^2+|\chi_+(\infty)|^2}, \label{alphabeta1}\\
\alpha_2 = \frac{2\chi_+(0)\varphi_+(0)}{|\varphi_+(0)|^2+|\chi_+(0)|^2}, &
\beta_2 = \frac{|\varphi_+(0)|^2-|\chi_+(0)|^2}{|\varphi_+(0)|^2+|\chi_+(0)|^2}.
\end{eqnarray}

For a given $k\in [-\frac12,\frac12]$ we will now calculate $\chi_+$ and $\varphi_+$ at $0$ and $\infty$ by integration of the LP in the $(f,g)$-plane. The starting point of the integration is $(\frac12,\frac12)$, where the normalisation \eref{NormChi} defines  $\varphi=1=\chi$. We achieve $\lambda=\infty$ at $(-k,\frac12)$ and $\lambda=0$ at $(\frac12,k)$. Choosing the integration path along $g=\frac12$ and $f=\frac12$ respectively now leads to two major simplifications. First of all, the LP can be reduced to a single ODE in both cases, and secondly only values on the boundaries of the IVP are used. Therefore $\alpha_{1/2}$ and $\beta_{1/2}$ can be calculated from the initial values alone.

For $-f<k<g$ the value of $\lambda$ is purely imaginary, and its imaginary part is positive on the inner side of $\Gamma$. Hence the ODEs read
\begin{eqnarray}
\fl \left(\begin{array}{cc} \varphi_+ \\ \chi_+ \end{array}\right)_f = \left(\begin{array}{cc} A & {\rm i}\lambda^f A \\ {\rm i}\lambda^f \bar{A} & \bar{A} \end{array}\right) \left(\begin{array}{cc} \varphi_+ \\ \chi_+ \end{array}\right), \quad\; \lambda^f:=-\rmi\lambda=\sqrt{\frac{\frac12-k}{f+k}}>0, \quad\; g=\frac12; \label{ODE1} \\
\fl \left(\begin{array}{cc} \varphi_+ \\ \chi_+ \end{array}\right)_g = \left(\begin{array}{cc} B & -{\rm i}\lambda^g B \\ -{\rm i}\lambda^g \bar{B} & \bar{B} \end{array}\right) \left(\begin{array}{cc} \varphi_+ \\ \chi_+ \end{array}\right), \quad\; \lambda^g:={\rm i}/\lambda=\sqrt{\frac{\frac12+k}{g-k}}>0, \quad\; f=\frac12, \label{ODE2}
\end{eqnarray}
where \eref{ODE1} shall be integrated from $f=\frac12$ to $f=-k$ and \eref{ODE2} shall be integrated from $g=\frac12$ to $g=k$ respectively. This corresponds to step (i) of the scheme in \fref{Lösungsschema}.

\subsection{The boundary values of the jump matrix}

We define the boundary coefficients $A_{\rm b}, B_{\rm b}\in \mathbb{C}$ as well as their amplitudes and phases $\rho_{1/2},\phi_A,\phi_B\in \mathbb{R}$ by
\begin{eqnarray}
\fl A_{\rm b}:=\rho_1 \rme^{\rmi \phi_A}:=\lim_{(f,g)\rightarrow(\frac12,\frac12)} \left[\sqrt{\mbox{$\frac12$}-f}A\right], 
\quad
B_{\rm b}:=\rho_2 \rme^{\rmi \phi_B}:=\lim_{(f,g)\rightarrow(\frac12,\frac12)} \left[\sqrt{\mbox{$\frac12$}-g}B\right].
\end{eqnarray}
Considering \eref{ABdurchE} we can thus state the colliding wave conditions \eref{cwc} as
\begin{eqnarray}
\rho_1=\sqrt{\frac{k_1}{2}},\quad\quad
\rho_2=\sqrt{\frac{k_2}{2}}.
\end{eqnarray}
From the domain \eref{kWerte} of $k_{1/2}$ we get
\begin{eqnarray} \label{rhoGrenze}
\mbox{$\frac12$}\leq \rho_{1/2} <2^{-\frac12}.
\end{eqnarray}

In order to calculate the boundary values of the jump matrix, $J(\pm\lambda_1)$ and $J(\pm\lambda_2)$, we examine \eref{ODE1} for $k=-\frac12+\epsilon$. Substituting $f=\frac12-\delta$, the ODE system is, to leading order in $\delta$, given by
\begin{eqnarray}
\fl \left(\begin{array}{cc} \varphi_+ \\ \chi_+ \end{array}\right)_\delta = -\delta^{-\frac12}\left(\begin{array}{cc} A_{\rm b} & {\rm i}\lambda^f A_{\rm b} \\ {\rm i}\lambda^f \bar{A}_{\rm b} & \bar{A}_{\rm b} \end{array}\right) \left(\begin{array}{cc} \varphi_+ \\ \chi_+ \end{array}\right), \quad\; f=\frac12-\delta,g=\frac12, \label{ODEdelta}
\end{eqnarray}
which has to be integrated from $\delta=0$ to $\delta=\epsilon$. For $0<\delta<\epsilon\ll 1$ we have $\lambda^f\gg1$ and can reduce \eref{ODEdelta} in leading order to 
\begin{eqnarray}
(\varphi_+)_\delta=-\rmi [\delta(\epsilon-\delta)]^{-\frac12}A_{\rm b} \chi_+,\quad
(\chi_+)_\delta=-\rmi [\delta(\epsilon-\delta)]^{-\frac12}\bar{A}_{\rm b} \varphi_+.
\end{eqnarray}
Substituting $s=2 \arcsin(\sqrt{\delta/\epsilon})$ we obtain with $\chi_+=1=\varphi_+$ at $s=0$ the solution
\begin{eqnarray}
\left(\begin{array}{cc} \varphi_+ \\ \chi_+ \end{array}\right)=\left(\begin{array}{cc} \cos (|A_{\rm b}|s)-\frac{{\rm i} A_{\rm b}}{|A_{\rm b}|} \sin (|A_{\rm b}|s) \\  \cos (|A_{\rm b}|s)-\frac{{\rm i} |A_{\rm b}|}{A_{\rm b}} \sin (|A_{\rm b}|s) \end{array}\right).
\end{eqnarray}
From the value at $s=\pi$ corresponding to $\lambda=\infty$ we get using \eref{alphabeta1} in the trivial limit $\epsilon\to 0$ the boundary values of the jump matrix elements at $\pm\lambda_1$:
\begin{eqnarray}
\alpha_{1{\rm b}}:=\alpha_{1}(k=-\mbox{$\frac12$}) = \cos (2\pi \rho_1)-{\rm i} \cos(\phi_A) \sin (2 \pi  \rho_1), \label{alpha1R}\\
\beta_{1{\rm b}}:=\beta_{1}(k=-\mbox{$\frac12$}) = \sin(\phi_A) \sin (2 \pi \rho_1).  \label{alphabeta1R}
\end{eqnarray}
In the same way we can derive 
\begin{eqnarray}
\alpha_{2{\rm b}}:=\alpha_{2}(k=\mbox{$\frac12$}) = \cos (2\pi \rho_2)+{\rm i} \cos(\phi_B) \sin (2 \pi  \rho_2), \label{alpha2R}\\
\beta_{2{\rm b}}:=\beta_{2}(k=\mbox{$\frac12$}) = - \sin(\phi_B) \sin (2 \pi \rho_2). \label{alphabeta2R}
\end{eqnarray}
With relation \eref{rhoGrenze} the range of $\Re(\alpha_{1{\rm b}})$ and $\Re(\alpha_{2{\rm b}})$ is
\begin{eqnarray}
-1\leq\Re(\alpha_{1/2{\rm b}}) < \cos(\sqrt{2}\pi)<0. \label{RealphaDom}
\end{eqnarray}
In particular $\alpha_{1/2{\rm b}}=1$, which would be necessary for a continuous connection to the jump matrix $\mathbb{1}$ on a continued contour, is not consistent with the colliding wave conditions.
The equality in \eref{RealphaDom} is reached for impulsive waves:
\begin{eqnarray}
\alpha_{1/2{\rm b}}=-1 \quad\Leftrightarrow\quad \rho_{1/2}=\mbox{$\frac12$}
\quad\Leftrightarrow\quad k_{1/2}=\mbox{$\frac12$} \quad\Leftrightarrow\quad n_{1/2}=2.
\end{eqnarray}

\subsection{Integral equations for the RHP}

Within the representation \eref{Matrixlösung} the jump equation \eref{Sprung} is equivalent to the scalar jump equation
\begin{equation}
\varphi_+=\alpha\varphi_-+\beta\varphi_+(-\lambda). \label{SprungSkalar}
\end{equation}
Using the additive jump function $\mu(\lambda')$ we can express $\varphi$ as the Cauchy integral 
\begin{equation}
\varphi(\lambda)=1+\frac{1}{2\pi \rmi}\int_{\Gamma}\left(\frac{1}{\lambda'-\lambda}-\frac{1}{\lambda'+1}\right)\mu(\lambda')\rmd\lambda'.
\end{equation}
With the Cauchy principal value $\fint$ the inner and outer limit of an integral
\begin{equation}
I(\lambda)=\frac{1}{2\pi \rmi}\int_C\frac{\mu(\lambda')\rmd\lambda'}{\lambda'-\lambda}
\end{equation}
over a contour $C$ through $\lambda$ can be represented as
\begin{equation*}
I_+(\lambda)=\frac{1}{2\pi \rmi}\fint_C\frac{\mu(\lambda')\rmd\lambda'}{\lambda'-\lambda}+\frac{1}{2}\mu(\lambda),\quad 
I_-(\lambda)=\frac{1}{2\pi \rmi}\fint_C\frac{\mu(\lambda')\rmd\lambda'}{\lambda'-\lambda}-\frac{1}{2}\mu(\lambda). 
\end{equation*}
Insertion into \eref{SprungSkalar} yields with $\mu_{1/2}:=\mu|_{\Gamma_{1/2}}$ and $F(\lambda,\lambda'):=(\lambda'-\lambda)^{-1}-(\lambda'+1)^{-1}$ for $\lambda\in\Gamma_1$ the integral equation
\begin{eqnarray} 
\fl 1+\frac{1}{2\pi \rmi}\fint_{\Gamma_1}F(\lambda,\lambda')\mu_1(\lambda'){\rm d}\lambda'+\frac{1}{2}\mu_1(\lambda)+\frac{1}{2\pi \rmi}\int_{\Gamma_2}F(\lambda,\lambda') \mu_2(\lambda'){\rm d}\lambda' \label{IntGl} \nonumber\\ \fl
\quad\quad =\alpha\left[1+\frac{1}{2\pi \rmi}\fint_{\Gamma_1}F(\lambda,\lambda') \mu_1(\lambda'){\rm d}\lambda'-\frac{1}{2}\mu_1(\lambda)+\frac{1}{2\pi \rmi}\int_{\Gamma_2}F(\lambda,\lambda') \mu_2(\lambda'){\rm d}\lambda'\right] \\ \fl 
\quad\quad+\beta\left[1+\frac{1}{2\pi \rmi}\fint_{\Gamma_1}F(-\lambda,\lambda') \mu_1(\lambda'){\rm d}\lambda'+\frac{1}{2}\mu_1(-\lambda)+\frac{1}{2\pi \rmi}\int_{\Gamma_2}F(-\lambda,\lambda') \mu_2(\lambda'){\rm d}\lambda'\right]  \nonumber
\end{eqnarray}
and a similar relation for $\lambda\in\Gamma_2$. These integral equations may be solved analytically for some special cases. In general they can be evaluated by an expansion in Chebyshev polynomials, which corresponds to step (iia) in \fref{Lösungsschema}. But due to the discontinuities of the jump matrix $J$, the scalar solution has divergences at the contour endpoints which recur also in the additive jump functions $\mu_{1/2}$. Therefore such an expansion is much more challenging than for integral equations for  regular  functions. Hence the transformation to the cRHP is not only necessary to prove the existence of RHP solutions fulfilling the holomorphcity conditions \eref{holoEigensch}, but also to obtain integral equations with better properties for numerical treatment.

\section{Transformation to a continuous Riemann-Hilbert problem}

\subsection{Concept of transformation}

The transformation to a continuous Riemann-Hilbert problem is inspired by a recipe described by Vekua in \cite{Vekua1967}, where a jump matrix discontinuity is removed through multiplication with an appropriate branch cut perpendicular to the contour. In our RHP we are facing 4 discontinuities at the endpoints of the partial contours $\Gamma_{1/2}$. We can simultaneously remove the two discontinuities at the endpoints of a single partial contour using the functions $L_{1/2}^{\rho_{1/2}}$ and $L_{1/2}^{\rho_{1/2}-1}$ featuring a branch cut along $\Gamma_{1/2}$. They contain the fractions
\begin{eqnarray}
L_1:=\frac{\lambda_1+\lambda}{\lambda_1-\lambda},\quad 
L_2:=\frac{\lambda+\lambda_2}{\lambda-\lambda_2}.
\end{eqnarray}
We use $L_{1/2}^{\rho_{1/2}}$ and $L_{1/2}^{\rho_{1/2}-1}$ as well as their inverses as functions only in the $\lambda$-sheet with real value at $\lambda=1$, where we regard them as having a jump on the contour $\Gamma_{1/2}$. The inner and outer limits at the contour are
\begin{eqnarray}
(L_1^{\rho_1})_+=e^{\pi i \rho_1}|L_1^{\rho_1}|, \quad (L_1^{\rho_1})_-=e^{-\pi i \rho_1}|L_1^{\rho_1}|, \quad\quad
\lambda\in\Gamma_1,\label{L1innenAussen} \\
(L_2^{\rho_2})_+=e^{-\pi i \rho_2}|L_2^{\rho_2}|, \quad (L_2^{\rho_2})_-=e^{\pi i \rho_2}|L_2^{\rho_2}|, \quad\quad
\lambda\in\Gamma_2,\label{L2innenAussen} 
\end{eqnarray}
and analogous for $L_{1/2}^{\rho_{1/2}-1}$. This implies $L_{1/2}^{\rho_{1/2}}(-\lambda)=L_{1/2}^{-\rho_{1/2}}$. For technical reasons we restrict our derivation of the cRHP to non-impulsive waves by demanding
\begin{eqnarray} \label{rhoGrenze2}
\mbox{$\frac12$}<\rho_{1/2}<2^{-\frac12}
\end{eqnarray}
and hence excluding the case $\rho_1=\frac12\vee\rho_2=\frac12$, where $L_{1/2}^{\rho_{1/2}}$ and $L_{1/2}^{\rho_{1/2}-1}$ become the inverse of each other. 

As in \cite{Vekua1967} we demand the jump matrix $J$ of the initial RHP to be Lipschitz continuous at the endpoints of $\Gamma$. Thus we can state for later reference
\begin{equation}
\lim_{\lambda\to\lambda_{1/2}}(\lambda-\lambda_{1/2})^x(J-J(\lambda_{1/2}))=0
\quad{\rm for}\quad |x|<1. \label{Stetigkeit}
\end{equation}

\subsection{The extended Riemann-Hilbert problem}

We introduce the extended RHP (eRHP) 
\begin{eqnarray}
\Phi_+=\Phi_- G \label{Sprungg}
\end{eqnarray}
with the slightly modified `generalised jump matrix'
\begin{eqnarray}
G=\left(\begin{array}{cc}
\alpha & \gamma+\beta \\ \gamma-\beta & \bar{\alpha} \label{gForm}
\end{array}\right), \quad \gamma,\beta\in\mathbb{R}
\end{eqnarray} 
defined on the whole real $\lambda$-axis (denoted by $\Gamma_\Re$) and featuring the properties
\begin{eqnarray}
\det G =\alpha\bar{\alpha} +\beta^2-\gamma^2=1,\quad\alpha(-\lambda)=\alpha,\beta(-\lambda)=\beta,\gamma(-\lambda)=-\gamma. \label{gStruktur}
\end{eqnarray}
For $\gamma\neq 0$ the eRHP jump matrix $G$ is neither unitary nor symmetric in $\lambda$ any more. The eRHP jump induced by the RHP jump matrix $J$ is denoted as
\begin{equation}
G_J:=\cases{ J           & $\lambda\in\Gamma$, \\
             \mathbb{1}  & else. }
\end{equation}
In the following subsections we will describe transformations like $G\to G'$ by expressing the new jump functions $\alpha'$, $\beta'$ and $\gamma'$ in terms of the old ones. The relations \eref{gStruktur} remain valid in all cases. The effect of the transformations leading to the cRHP are illustrated in \fref{fig:VisTrafo}.

Note that already the special case $\alpha|_{\Gamma_2}=\rm{const}$, $\alpha|_{\Gamma_\Re\backslash\Gamma_2}=1$, $\beta=0=\gamma$ features two independent scalar solutions $\varphi=L_2^{\rho_2}$ and $\varphi=L_2^{\rho_2-1}$. This ambiguity is connected with the discontinuities of the jump function $\alpha$ at $\pm\lambda_2$. Within the transformation to the cRHP, these ambiguities arise 
at each partial contour in the shape of two different ways of removing the discontinuities.  

For clarity of notation, we treat the eRHP without normalisation. Out of the scalar RHP solution $\varphi$ the Ernst potential $E=\varphi(1)/\varphi(-1)$ with the right normalisation \eref{NormErnst} can be easily derived afterwards.

\subsection{Rotation transformation}

We define a rotation transformation, which converts \eref{Sprungg} to $\Phi'_+=\Phi'_- G'$ by
\begin{eqnarray}
\Phi'=\Phi R_\delta, \quad G'=R_\delta^{-1}G R_\delta, \quad\quad R_\delta=\left(\begin{array}{cc}
\cos{\delta} & \rmi \sin{\delta} \\
\rmi \sin{\delta} & \cos{\delta} 
\end{array}\right).\label{x-y-Rotation}
\end{eqnarray}
The scalar solution and the Ernst potential transform as 
\begin{eqnarray}
\varphi'=\cos{\delta}\varphi-\rmi\sin{\delta}\varphi(-\lambda), \quad 
E'=\cos{\delta}E-\rmi\sin{\delta}. \label{eq:xy-RotationSkalar}
\end{eqnarray}

Note that if $\Phi$ was normalised and we normalise $\Phi'$ according to \eref{NormRHP} by
\begin{eqnarray}
\Phi''=T\Phi', \quad T=\rm{diag}\lbrace
\frac{1}{\cos{\delta}-\rmi\sin{\delta}E} , \frac{1}{\cos{\delta}+\rmi\sin{\delta}\bar{E}}\rbrace,
\end{eqnarray}
we get 
\begin{equation} \label{Estrichstrich}
E''=\frac{\cos{\delta}E-\rmi\sin{\delta}}{\cos{\delta}-\rmi\sin{\delta}E}. 
\end{equation}
This is the corresponding Ernst potential for a metric of the form \eref{metric} after a clockwise rotation of the $x$-$y$-plane by an angle $\delta$,
 \begin{equation}
\left(\begin{array}{c} x'' \\ y'' \end{array}\right) =
\left(\begin{array}{cc}
\cos{\delta} & -\sin{\delta} \\
 \sin{\delta} & \cos{\delta}
\end{array}\right)
\left(\begin{array}{c} x \\ y \end{array}\right).
\end{equation}
Secondly, \eref{Estrichstrich} is exactly the result of an `Ehlers transformation' $E''=E/(1-\rmi\tan{\delta}E)$ with subsequent normalisation in virtue of \eref{NormErnst}.  

The jump functions transform under \eref{x-y-Rotation} as
\begin{eqnarray}\begin{array}{ll}
\Im (\alpha')=\cos(2\delta)\Im (\alpha)+\sin(2\delta)\beta, \quad\quad\quad&
\Re (\alpha')=\Re (\alpha'), \\
\beta'=-\sin(2\delta)\Im (\alpha)+\cos(2\delta)\beta, &
\gamma'=\gamma.
\end{array}\end{eqnarray}
Starting with the induced eRHP jump $G_J$, the boundary values \eref{alpha1R}-\eref{alphabeta2R} of the RHP jump functions transform under the rotation transformation \eref{x-y-Rotation} to
\begin{eqnarray}
\fl\alpha'_{1{\rm b}} &= \cos (2\pi \rho_1)-{\rm i} \cos(\phi_A+2\delta) \sin (2 \pi  \rho_1), \quad
&\beta'_{1{\rm b}} =  \sin(\phi_A+2\delta) \sin (2 \pi \rho_1),  \\
\fl\alpha'_{2{\rm b}} &= \cos (2\pi \rho_2)+{\rm i} \cos(\phi_B+2\delta) \sin (2 \pi  \rho_2), \quad
&\beta'_{2{\rm b}} = - \sin(\phi_B+2\delta) \sin (2 \pi \rho_2).
\end{eqnarray}
Thus the clockwise coordinate rotation in the $x$-$y$-plane by an angle $\delta$ corresponds to a counterclockwise rotation of $A_{\rm b}$ and $B_{\rm b}$ in the complex plane by an angle $2\delta$.
If the initial values imply $\phi_A-\phi_B=n \pi, n\in \mathbb{R}$, then the RHP jump matrix can be diagonalized at all 4 contour endpoints simultaneously, which leads to tremendous simplifications in the transition to a continuous RHP. We will call this case `initially collinearly polarised GW'.

By convention we choose for the diagonalisation of $J_{1/2{\rm b}}$ the rotation matrices $R_{\delta_1}$ and $R_{\delta_2}$ with
\begin{eqnarray}
\delta_1:=(\pi-\phi_A)/2,\quad \delta_2:=(\pi-\phi_B)/2.
\end{eqnarray}
With these transformations we can achieve 
\begin{eqnarray}
\fl G'|_{\pm(\lambda_1+0)}&=\left(\begin{array}{cc} e^{2\pi \rmi \rho_1} & 0 \\  0 & e^{-2\pi \rmi \rho_1} \end{array}\right) \quad {\rm or}\quad
G'|_{\pm(\lambda_2-0)}&=\left(\begin{array}{cc} e^{-2\pi \rmi \rho_2} & 0 \\  0 & e^{2\pi \rmi \rho_2} \end{array}\right). \label{gstrichRand}
\end{eqnarray}
From now on we use our freedom of a rotation in the $x$-$y$-plane to choose coordinates so that the jump matrix is initially diagonal at $\pm\lambda_2$, i.e. $G_J|_{\pm(\lambda_2-0)}={\rm diag}(e^{-2\pi \rmi \rho_2} , e^{2\pi \rmi \rho_2})$.

\subsection{Singularity transformation} \label{SingTrafo}

We define a singularity transformation, which converts \eref{Sprungg} to $\tilde{\Phi}_+=\tilde{\Phi}_-\tilde{G}$ by
\begin{eqnarray}
\tilde{\Phi}=\Phi S^K_{1/2}, \quad \tilde{G}^K=(S^K_{1/2-})^{-1} G S^K_{1/2+}. \label{SiTrafo}
\end{eqnarray}
Herein $K$ is an index which takes the values `$\rm e$' and `$\rm o$' designating the two possibilities of using either an even or an odd singularity transformation matrix,
\begin{eqnarray}
S_{1/2}^{\rm e}:=\left(\begin{array}{cc}
L_{1/2}^{1-\rho_{1/2}} & 0 \\
0 & L_{1/2}^{\rho_{1/2}-1} 
\end{array}\right) \quad\rm{or}\quad
S_{1/2}^{\rm o}:=\left(\begin{array}{cc}
L_{1/2}^{1-\rho_{1/2}} & 0 \\
0 & L_{1/2}^{\rho_{1/2}} 
\end{array}\right). \label{Yeo}
\end{eqnarray}
Evaluation of the inner and outer limits similar to \eref{L2innenAussen} leads to  
\begin{eqnarray} 
\tilde{\gamma}^K=\varepsilon^K_{1/2} \frac{1}{2}\left[(|L_{1/2}|^{x_{1/2}^K}+|L_{1/2}|^{-x_{1/2}^K})\gamma +  (|L_{1/2}|^{x_{1/2}^K}-|L_{1/2}|^{-x_{1/2}^K}) \beta \right],     \\
\tilde{\beta}^K=\varepsilon^K_{1/2} \frac{1}{2}\left[(|L_{1/2}|^{x_{1/2}^K}-|L_{1/2}|^{-x_{1/2}^K})\gamma +  (|L_{1/2}|^{x_{1/2}^K}+|L_{1/2}|^{-x_{1/2}^K}) \beta \right],
\\
\tilde{\alpha}^K=
\cases{  \rme^{\mp 2\pi \rmi\rho_{1/2}}\alpha & $\!\!\lambda\in\Gamma_{1/2}$, \\ \alpha & $\!\!$ else}
\quad \mbox{(`-' associated with index `1')}
\end{eqnarray}
with $\varepsilon^{\rm e}_{1/2}=1=\varepsilon^{\rm o}_{1/2}|_{\Gamma_\Re / \Gamma_{1/2}}$, $\varepsilon^{\rm o}_{1/2}|_{\Gamma_{1/2}}=-1$, $x_{1/2}^{\rm e}=2\rho_{1/2}-2$ and $x_{1/2}^{\rm o}=2\rho_{1/2}-1$.\\[-0.3cm]

Considering \eref{Stetigkeit}, due to $|x_{1/2}^K|<1$ we get by applying the singularity transformation to a RHP jump matrix diagonalised at $\pm\lambda_2$:
\begin{equation}
\tilde{G}_J^K:=(S^K_{2-})^{-1}G_JS^K_{2+}, \quad \tilde{G}_J^K(\pm\lambda_2)=\mathbb{1}.
\end{equation}
The jump matrix $\tilde{G}_J^K$ is continuous at $\pm\lambda_2$, but not necessarily Lipschitz continuous, whereas at $\pm\lambda_1$ the jump matrix is still Lipschitz continuous. However, $\tilde{G}_J^K$ is no longer unitary, so another type of transformation is necessary to restore the unitarity of the jump matrix at $\pm\lambda_1$ in order to diagonalise it by a rotation transformation and make it continuous by a singularity transformation afterwards.

\subsection{Unitarisation transformation}

We define a unitarisation transformation, which converts \eref{Sprungg} to $\hat{\Phi}_+=\hat{\Phi}_-\hat{G}$ by
\begin{eqnarray}
\fl\hat{\Phi}=\Phi U^K,\quad\hat{G}=(U^{K}_-)^{-1} G U^K_+,
\quad U^K:=\left(\begin{array}{cc} w^K\Lambda^K & 0 \\ 0 & (w^K\Lambda^K)^{-1} \end{array}\right),
\end{eqnarray}
where the constituents of the unitarisation matrix $U^K$ are defined as
\begin{eqnarray}
w^K:=\left({\rm sign}\left[\Lambda^K(\lambda_1)\right]\right)^{-1}, \quad \Lambda^K:=
\cases{  \frac{\lambda+\lambda^K_u}{\lambda-\bar{\lambda}^K_u} & $\Im(\lambda)>0$, \\ \frac{\lambda+\bar{\lambda}^K_u}{\lambda-\lambda^K_u} & $\Im(\lambda)<0$.}
\end{eqnarray}
Therein ${\rm sign}(a):=e^{i\arg(a)}$ is the complex generalisation of the sign function. The phase factor $w^K$ is constant in each half-plane and compensates the phase of $\Lambda^K$ in $\pm\lambda_1$. Similar to $L_{1/2}$, the functions $w^K$ and $\Lambda^K$ obey  $\bar{w}^K(\bar{\lambda})=w^K(\lambda)=1/w^K(-\lambda)$ and $\bar{\Lambda}^K(\bar{\lambda})=\Lambda^K(\lambda)=1/\Lambda^K(-\lambda)$. The jump functions are mapped to
\begin{eqnarray} \fl\begin{array}{l}
\hat{\gamma}= \frac{1}{2}\left[(|\Lambda^K|^{-2}+|\Lambda^K|^2)\gamma +  (|\Lambda^K|^{-2}-|\Lambda^K|^2) \beta \right],     \\
\hat{\beta}= \frac{1}{2}\left[(|\Lambda^K|^2-|\Lambda^K|^{-2})\gamma +  (|\Lambda^K|^{-2}+|\Lambda^K|^2) \beta \right],
\end{array}\quad\;\;\label{eq:gammaBetaSchlange} 
\hat{\alpha}=(w^K_+)^2{\,\rm sign}^2(\Lambda^K_+)\alpha.
\end{eqnarray}

If we choose $\lambda^K_u$ so that $|\Lambda^K(\lambda_1)|^2=\left(L_2(\lambda_1)\right)^{x_{2}^K}$, the unitarisation transformation applied after removing the discontinuities at $\pm\lambda_2$ yields 
\begin{equation}
\hat{G}_J^K:=(U^{K}_-)^{-1}\tilde{G}_J^KU^K_+, \quad \hat{G}_J^K|_{\pm(\lambda_1-0)}=\mathbb{1}, \quad \hat{G}_J^K|_{\pm(\lambda_1+0)}=J(\pm\lambda_1).
\end{equation}
Hence the unitarisation transformation reproduces the initial settings at $\pm\lambda_1$ with $\hat{G}_J^K$ still Lipschitz continuous at these points. Furthermore $\hat{G}_J^K\neq\mathbb{1}$ almost everywhere on $\Gamma_\Re$ and so the matrix solution $\hat{\Phi}$ is no longer described by a single expression for both sides of the contour. With $\Im(\lambda^K_u)>0$ we assure that $\Lambda^K$ has neither zeros nor poles.

\subsection{The full transformation formula}

We can now diagonalise the jump matrix $\hat{G}_J^K$ at $\pm\lambda_1$ by the  rotation transformation 
\begin{equation}
{G'}_J^K:=R_{\delta_1}^{-1}\hat{G}_J^KR_{\delta_1}
\end{equation}
and remove the discontinuities there by a singularity transformation using $S^I_{1}$ analogous to the procedure at $\pm\lambda_2$. In summary, we get the cRHP
\begin{equation}
\Omega^{IK}_+=\Omega^{IK}_- G_{\rm c}^{IK}, \quad\quad G_{\rm c}^{IK}\in C^0\label{cSprung}
\end{equation}
from the eRHP \eref{Sprungg} by applying the transformation
\begin{eqnarray}
\Omega^{IK}:=\Phi S^K_2 U^K R_{\delta_1} S^I_1, \label{cRHPsolution} \\
G_{\rm c}^{IK}:=(S^I_{1-})^{-1} R_{\delta_1}^{-1} (U^K_-)^{-1} (S^K_{2-})^{-1} G_J S^K_{2+} U^K_+ R_{\delta_1} S^I_{1+}. \label{cRHPjump}
\end{eqnarray}
The jump matrix $G_{\rm c}^{IK}$ of the cRHP depends in contrast to $G_J$ on the coordinates $f$ and $g$. This is an interesting similarity to the treatment of Alekseev and Griffiths \cite{Alekseev_Griffiths2004}, where the non-analytic behaviour of the solution at the wavefronts is handled by `dynamical' monodromy data and generalised integral evolution equations.

\subsection{The degree of the solution row vectors}

We fix a point $\lambda_p$ on the imaginary axis of the $\lambda$-plane (one may think of $\lambda_p=\rmi$) and define a $\lambda_p$-regular function as a function which is only allowed to have poles or zeros in $\lambda_p$. Furthermore, we define the degree of a $\lambda_p$-regular function $f(\lambda)$ as
\begin{eqnarray*}
\mbox{degree of $f(\lambda)$ } 
&:=\cases{ n & $f(\lambda)$ has pole of order $n$ in $\lambda_p$, \\
		   0 & $0\neq f(\lambda_p)\neq\infty$,  \\
		  -n & $f(\lambda)$ has zero of order $n$ in $\lambda_p$.} 
\end{eqnarray*}
The same definition applies to matrices, keeping in mind that a matrix has a pole where one element has pole and a zero where all elements have a zero.

According to \cite{Vekua1967} (where a finite contour with $\lambda_p=\infty$ is discussed), a two-dimensional cRHP has a fundamental matrix $\Omega^{IK}=(\Omega^{IK}_1;\Omega^{IK}_2)$ characterized by the $\lambda_p$-regular and linearly independent solution row vectors $\Omega^{IK}_1$ and $\Omega^{IK}_2$ having minimal degree $\varkappa_1$ and $\varkappa_2$, respectively. From the fundamental matrix all solution vectors can be constructed as linear combinations. It is shown in \cite{Vekua1967} to have the following two properties:
\begin{eqnarray}
\det\Omega^{IK}\neq 0\;\forall \lambda\neq\lambda_p; \\
0<(\lambda-\lambda_p)^{\varkappa_1+\varkappa_2}\det\Omega^{IK}(\lambda_p)<\infty.
\end{eqnarray}
Thus we can conclude that $\det\Omega^{IK}$ is $\lambda_p$-regular with degree $\varkappa_D=\varkappa_1+\varkappa_2$. Furthermore, due to $\det G_{\rm c}^{IK}=\mathbb{1}$ the determinant of the cRHP solution $\Omega^{IK}$ has no jump on $\Gamma_\Re$. In consequence $\varkappa_D=0=\varkappa_1+\varkappa_2$.

\subsection{The fundamental matrix of the cRHP}
 
From \eref{gForm} with \eref{gStruktur} we can derive on $\Gamma_\Re$ the identities
\begin{eqnarray}
\mathbb{1}=\sigma_1 G_{\rm c}^{IK}(-\lambda)\sigma_1 G_{\rm c}^{IK},\quad\quad 
\mathbb{1}=\sigma_3 \bar{G}_{\rm c}^{IK}\sigma_3 G_{\rm c}^{IK}.
\end{eqnarray}
Inserting \eref{cSprung} for $G_{\rm c}^{IK}(-\lambda)$ and $\bar{G}_{\rm c}^{IK}$ we see that $\Omega^{IK}(-\lambda)\sigma_1$ and $\bar{\Omega}^{IK}(\bar{\lambda})\sigma_3$ fulfil the same jump equation as $\Omega^{IK}$. This statement holds already for row vectors. Therefore, from an arbitrary $\lambda_p$-regular solution row vector $\Omega_1$ of the cRHP with minimal degree $\varkappa_1$ we can construct another $\lambda_p$-regular solution row vector with degree $\varkappa_1$ represented by a single scalar function $\vartheta^{IK}$,
\begin{equation}
\fl\Theta^{IK}_1:=\left(\vartheta^{IK},-L_p^{\varkappa_1}\vartheta^{IK}(-\lambda)\right)
:=\Omega^{IK}_1-L_p^{\varkappa_1}\Omega^{IK}_1(-\lambda)\sigma_1, \quad
L_p:=\frac{\lambda_p+\lambda}{\lambda_p-\lambda}.
\end{equation}
Within the summation, no new zeros can arise because $\Omega_1$ has already minimal degree. From defining the matrix $\Theta^{IK}:=\left(\Theta^{IK}_1;L_p^{\varkappa_1}\bar{\Theta}^{IK}_1(\bar{\lambda})\sigma_3\right)$ and calculating
\begin{equation}
\det\Theta^{IK}(0)=|\vartheta^{IK}_{+}(0)|^2+|\vartheta^{IK}_{-}(0)|^2\neq 0 
\end{equation}
we see that $\Theta^{IK}_1$ and $L_p^{\varkappa_1}\bar{\Theta}^{IK}_1(\bar{\lambda})\sigma_3$ are both linearly independent $\lambda_p$-regular solution vectors with degree $\varkappa_1$. Thus $\varkappa_1=\varkappa_2=0$ and we can write the fundamental matrix of the cRHP in normal form with the scalar function $\vartheta^{IK}$:
\begin{eqnarray} 
\Theta^{IK}=\left(\begin{array}{cc} \vartheta^{IK} & -\vartheta^{IK}(-\lambda) \\  \bar{\vartheta}^{IK}(\bar{\lambda})& \bar{\vartheta}^{IK}(-\bar{\lambda}) \end{array}\right).
\label{MatrixForm}
\end{eqnarray}
The row vectors constituting the fundamental matrix feature neither zeros nor poles. The jump equation $\Theta^{IK}_+=\Theta^{IK}_-G_{\rm c}^{IK}$ is equivalent to the single scalar jump equation
\begin{eqnarray}
\vartheta^{IK}_+=\alpha_c \vartheta^{IK}_-+(\beta_c-\gamma_c)\vartheta^{IK}_+(-\lambda). \label{Sprungtheta}
\end{eqnarray}

\subsection{The normal form solution of the RHP}

We now gradually revert the transformation \eref{cRHPsolution} and ensure in each step the normal form of the matrix solution.
In the `even' case the first partial inverse transformation ${\Theta'}^{{\rm e}K}:=\Theta^{{\rm e}K}(S_1^{\rm e})^{-1}$ yields directly a matrix in normal form with 
\begin{equation}
{\vartheta'}^{{\rm e}K}=L_1^{\rho_1-1}\vartheta^{{\rm e}K}. \label{S1eRücktrafo}
\end{equation}
In the `odd' case, $\Theta^{{\rm o}K}(S_1^{\rm o})^{-1}$ is not in normal form, but we can obtain a solution ${\Theta'}^{{\rm o}K}$ to the jump equation ${\Theta'}_+^{{\rm o}K}={\Theta'}_-^{{\rm o}K}{G'_J}^{\rm o}$ in normal form with the scalar function 
\begin{equation}
{\vartheta'}^{{\rm o}K}=L_1^{\rho_1-1}(1+L_1)\vartheta^{{\rm o}K} \label{S1oRücktrafo}
\end{equation}
via the linear combination ${\Theta'}^{{\rm o}K}:=\Theta^{{\rm o}K}(S_1^{\rm o})^{-1}-\sigma_3\Theta^{{\rm o}K}(-\lambda)(S_1^{\rm o})^{-1}(-\lambda)\sigma_1$. 
Note that if we had defined the odd transformation by the alternative matrix
\begin{eqnarray}
S_{1}^{\rm o\ast}:=\left(\begin{array}{cc}
L_1^{-\rho_1} & 0 \\
0 & L_1^{\rho_1-1} 
\end{array}\right)=L_1^{-1} S_{1}^{\rm o}
\end{eqnarray}
instead of $S_{1}^{\rm o}$, we would after inverse transformation end up with the same normal form solution corresponding to \eref{S1oRücktrafo}. 

The second partial inverse transformation $\hat{\Theta}^{IK}={\Theta'}^{IK}R^{-1}_{\delta_1}$ yields directly a matrix $\hat{\Theta}^{IK}$ in normal form with
\begin{eqnarray}
\hat{\vartheta}^{IK}=\cos \delta_1 {\vartheta'}^{IK} +\rmi \sin \delta_1 {\vartheta'}^{IK}(-\lambda).
\end{eqnarray}
Likewise the inverse transformation $\tilde{\Theta}^{IK}=\hat{\Theta}^{IK}(U^K)^{-1}$ yields a matrix ${\tilde{\Theta}}^{IK}$ in normal form with
\begin{eqnarray}
\tilde{\vartheta}^{IK}=(w^K\Lambda^K)^{-1}\hat{\vartheta}^{IK}.
\end{eqnarray}
At last, after inverse transformation with $(S_2^{K})^{-1}$ we obtain, analogous to \eref{S1eRücktrafo} and \eref{S1oRücktrafo}, a solution $\Phi^{IK}$ to the initial RHP in normal form with one of the scalar functions
\begin{eqnarray}
\varphi^{I \rm{e}}=L_2^{\rho_2-1}\tilde{\vartheta}^{I\rm{e}} \quad \mbox{or}\quad
\varphi^{I \rm{o}}=L_2^{\rho_2-1}(1+L_2)\tilde{\vartheta}^{I\rm{o}}.
\end{eqnarray}
In summary, via the cRHP we obtain the 4 matrix solutions $\Phi^{\rm ee}$, $\Phi^{\rm oe}$, $\Phi^{\rm eo}$ and $\Phi^{\rm oo}$ in normal form with the scalar functions given in terms of the cRHP solutions $\vartheta^{IK}$ as
\begin{eqnarray}
\fl\varphi^{\rm ee}(\lambda)=L_2^{\rho_2-1}(w^{\rm e}\Lambda^{\rm e})^{-1}\left[\cos\delta_1 L_1^{\rho_1-1}\vartheta^{\rm ee}+\rmi\sin\delta_1 L_1^{1-\rho_1}\vartheta^{\rm ee}(-\lambda)\right],
\nonumber\\ \fl
\varphi^{\rm oe}(\lambda)=L_2^{\rho_2-1}(w^{\rm e}\Lambda^{\rm e})^{-1}\left[\cos\delta_1 L_1^{\rho_1-1}(1+L_1)\vartheta^{\rm oe}+\rmi\sin\delta_1 L_1^{1-\rho_1}(1+L_1^{-1})\vartheta^{\rm oe}(-\lambda)\right], \nonumber\\ \fl
\varphi^{\rm eo}(\lambda)=L_2^{\rho_2-1}(1+L_2)(w^{\rm o}\Lambda^{\rm o})^{-1}\left[\cos\delta_1 L_1^{\rho_1-1}\vartheta^{\rm eo}+\rmi\sin\delta_1 L_1^{1-\rho_1}\vartheta^{\rm eo}(-\lambda)\right], \label{Chieeoo}\\ 
\fl\varphi^{\rm oo}(\lambda)=L_2^{\rho_2-1}(1+L_2)(w^{\rm o}\Lambda^{\rm o})^{-1} \nonumber\\ \cdot\left[\cos\delta_1 L_1^{\rho_1-1}(1+L_1)\vartheta^{\rm oo}+\rmi\sin\delta_1 L_1^{1-\rho_1}(1+L_1^{-1})\vartheta^{\rm oo}(-\lambda)\right]. \nonumber
\end{eqnarray}
The solution of the four scalar jump equations \eref{Sprungtheta} and the construction of these RHP solutions is subsumed in step (iia) of the solution scheme in \fref{Lösungsschema}.

\section{Regularity conditions for solutions of the linear problem}

\subsection{Construction of the solution to the linear problem}

Out of the RHP solutions \eref{Chieeoo} we construct the LP solution as linear combination
\begin{eqnarray}
\Phi^{\rm LP}=\Phi^{\rm oo}+{\rm diag}(p,\bar{p})\Phi^{\rm eo}+{\rm diag}(q,\bar{q})\Phi^{\rm oe}+{\rm diag}(r,\bar{r})\Phi^{\rm ee}, \\
\varphi^{\rm LP}=\varphi^{\rm oo}+p\varphi^{\rm eo}+q\varphi^{\rm oe}+r\varphi^{\rm ee}
\end{eqnarray}
and the LP matrices 
\begin{equation}
U=\Phi^{\rm LP}_f(\Phi^{\rm LP})^{-1}, \quad V=\Phi^{\rm LP}_g(\Phi^{\rm LP})^{-1}. \label{LPMat}
\end{equation}
The $\varphi$-coefficients $p$, $q$ and $r$ are functions of the coordinates $f$ and $g$ and have to be arranged to make $U$ holomorphic in $\mathbb{C}_{\lambda}\backslash\{\infty\}$ and $ V$ holomorphic in $\mathbb{C}_{\lambda}\backslash\{0\}$.
These are the generic regularity conditions \eref{holoEigensch}, which will be specified now. We start with an investigation of $\det\Phi^{\rm LP}$ before we examine $U$ and $V$ directly.

Due to the property \eref{DetPhi} of the LP, $\det\Phi^{\rm LP}$ has to be independent of $\lambda$. Because of $\det G_J=\mathbb{1}$ and the absence of poles in $\vartheta^{IK}$, this is the case if $\det\Phi^{\rm LP}$ has no poles in $\pm\lambda_{1/2}$. Since $\det\Phi^{\rm LP}(-\lambda)=\det\Phi^{\rm LP}$ because of the normal form, it is sufficient to examine the points $\lambda_{1/2}$.

\subsection{Regularity condition for $\det \Phi^{\rm LP}$ at $\lambda_2$}

In order to derive a first condition for the $\varphi$-coefficients from the $\lambda$-independence of $\det\Phi^{\rm LP}$, we collect the constituents of $\varphi^{\rm LP}$ regular in $\lambda_2$ as
\begin{eqnarray}
\fl\psi_2^{\rm e}=(w^{\rm e}\Lambda^{\rm e})^{-1} \lbrace\cos\delta_1 L_1^{\rho_1-1}\left[(1+L_1)\vartheta^{\rm oe}+q^{-1}r\vartheta^{\rm ee}\right]  \nonumber\\
+\rmi\sin\delta_1 L_1^{1-\rho_1}\left[(1+L_1^{-1})\vartheta^{\rm oe}(-\lambda)+q^{-1}r\vartheta^{\rm ee}(-\lambda)\right]\rbrace , 
 \\
\fl\psi_2^{\rm o}=(w^{\rm o}\Lambda^{\rm o})^{-1} \lbrace\cos\delta_1 L_1^{\rho_1-1}\left[(1+L_1)\vartheta^{\rm oo}+p\vartheta^{\rm eo}\right] \nonumber\\
+\rmi\sin\delta_1 L_1^{1-\rho_1}\left[(1+L_1^{-1})\vartheta^{\rm oo}(-\lambda)+p\vartheta^{\rm eo}(-\lambda)\right]\rbrace . 
\end{eqnarray}
The corresponding normal form matrices $\Psi_2^{\rm e}$ and $\Psi_2^{\rm o}$ are solutions of $\Psi_{2+}^I=\Psi_{2-}^I\tilde{G}_J^I$. Because of $\tilde{G}_J^I(\pm\lambda_2)=\mathbb{1}$ these scalar solutions have no jump at $\pm\lambda_2$:
\begin{equation}
\psi_{2+}^{I}(\pm\lambda_2)=\psi_{2-}^{I}(\pm\lambda_2). \label{psi2NoJump}
\end{equation}
The scalar LP solution can now be expressed as  
\begin{eqnarray}
\varphi^{\rm LP}(\lambda)&=L_2^{\rho_2-1}\left[(1+L_2)\psi_2^{\rm o}+q\psi_2^{\rm e}\right]. \label{psiugDarst}
\end{eqnarray}
The determinant $\det\Phi^{\rm LP}$ is regular (i.e. non-singular) in $\lambda_2$ if and only if the prefactor of $L_2$ in $\det\Phi^{\rm LP}$ vanishes at $\lambda_2$. Since we have \eref{psi2NoJump} this is equivalent to 
\begin{equation} \label{kappa2}
(\kappa_2+\bar{\kappa}_2)|\psi_{2}^{\rm o}(\lambda_2)|^2=0, \quad \kappa_2:=(\psi_2^{\rm o}(\lambda_2))^{-1} \left[\psi_2^{\rm o}(-\lambda_2)+q\psi_2^{\rm e}(-\lambda_2)\right].
\end{equation}

\subsection{Regularity condition for $\det \Phi^{\rm LP}$ at  $\lambda_1$} \label{s:detPhiLambda1}


We introduce for the prefactors of the scalar solutions $\varphi^{IK}$ the notation
\begin{eqnarray}
H^{\rm e}:=L_2^{\rho_2-1}(w^{\rm e}\Lambda^{\rm e})^{-1},\quad 
H^{\rm o}:=L_2^{\rho_2-1}(1+L_2)(w^{\rm o}\Lambda^{\rm o})^{-1}.
\end{eqnarray}
Due to the definition of $w^I$ and $\Lambda^I$ we have at $\pm\lambda_1$:
\begin{eqnarray}\fl\begin{array}{l}
H^{\rm e}(-\lambda_1)=1=H^{\rm e}(\lambda_1), \\
H^{\rm o}(-\lambda_1)=L_{12}^{-1/2}+L_{12}^{1/2}=H^{\rm o}(\lambda_1),
\end{array} \quad
L_{12}:=L_1(\lambda_2)=L_2(\lambda_1)=\frac{\lambda_1+\lambda_2}{\lambda_1-\lambda_2}. \label{L12}
\end{eqnarray}
In particular $H^{\rm e}$ and $H^{\rm o}$ have no jump at $\pm\lambda_1$. We define the following constituents of $\varphi$ regular in $\pm\lambda_1$ (with indices $p$ and $m$ abbreviating `plus' and `minus'):
\begin{equation}
\begin{array}{ll}
\psi_{1p}^{\rm e}:=H^{\rm o}\vartheta^{\rm eo}+\frac{r}{p} H^{\rm e}\vartheta^{\rm ee}, &\quad
\psi_{1m}^{\rm e}:=H^{\rm o}\vartheta^{\rm eo}(-\lambda)+\frac{r}{p} H^{\rm e}\vartheta^{\rm ee}(-\lambda), \\
\psi_{1p}^{\rm o}:=H^{\rm o}\vartheta^{\rm oo}+q H^{\rm e}\vartheta^{\rm oe}, &\quad
\psi_{1m}^{\rm o}:=H^{\rm o}\vartheta^{\rm oo}(-\lambda)+q H^{\rm e}\vartheta^{\rm oe}(-\lambda). \label{psiugpm}
\end{array}\end{equation}
Because of $G_{\rm c}^{IK}=\mathbb{1}$ the scalar cRHP solutions $\vartheta^{IK}$ have no jump in $\pm\lambda_1$ and using \eref{L12} we obtain
\begin{eqnarray}
\psi_{1m+}^I(\pm\lambda_1)=\psi_{1m-}^I(\pm\lambda_1)
=\psi_{1p+}^I(\mp\lambda_1)=\psi_{1p-}^I(\mp\lambda_1).\label{psipm}
\end{eqnarray}
Using $\psi_{1p/m}^I$ the scalar LP solution $\varphi^{\rm LP}$ can be expressed as
\begin{eqnarray}
\fl \varphi^{\rm LP}(\lambda)=\cos{\delta}L_1^{\rho_1-1}(1+L_1)\psi_{1p}^{\rm o} 
 +\rmi \sin{\delta}L_1^{1-\rho_1}(1+L_1^{-1})\psi_{1m}^{\rm o}\nonumber\\
+p\left[\cos{\delta}L_1^{\rho_1-1}\psi_{1p}^{\rm e}
 +\rmi\sin{\delta}L_1^{1-\rho_1}\psi_{1m}^{\rm e}\right].
 \label{chipsi1}
\end{eqnarray}
Considering \eref{psipm} the vanishing of the prefactor of $L_1$ in $\det\Phi^{\rm LP}$ at the point $\lambda_1$ can be shown to be equivalent to
\begin{equation} \label{kappa1}
\fl(\kappa_1+\bar{\kappa}_1)|\psi_{1p}^{\rm o}(\lambda_1)|^2=0, \quad \kappa_1:=(\psi_{1p}^{\rm o}(\lambda_1))^{-1} \left[\psi_{1p}^{\rm o}(-\lambda_1)+p\psi_{1p}^{\rm e}(-\lambda_1)\right].
\end{equation}
Unlike the situation at $\lambda_2$, during the calculation of $\det\Phi^{\rm LP}$ out of \eref{chipsi1}, in principle terms proportional to $L_1^{2\rho_1}$, $L_1^{2\rho_1-1}$ and $L_1^{2-2\rho_1}$ could occur. However, these terms with non-integer exponent are  associated with branch cuts, which have to lie on $\Gamma_\Re$ because $\vartheta^{IK}$ and the previously constructed transformations were continuous everywhere else in the $\lambda$-plane. But since $\det G_J=1$ the  determinant $\det\Phi^{\rm LP}$ has no jump on $\Gamma_\Re$ and such a branch cut is excluded. In summary, \eref{kappa2} and \eref{kappa1} are the conditions on the $\varphi^{\rm LP}$-coefficients $p$, $q$ and $r$ assuring that $\det\Phi^{\rm LP}$ does not depend on $\lambda$.

\subsection{Construction of the LP matrices $U$ and $V$}

From the normal form of $\Phi^{\rm LP}$ we derive
\begin{eqnarray}
\fl (\det\Phi^{\rm LP}) U  \label{DchiU}\\
\fl \qquad\qquad=\left(\begin{array}{cc}
\varphi^{\rm LP}_f(\lambda)\bar{\varphi}^{\rm LP}(-\bar{\lambda})+\varphi^{\rm LP}_f(-\lambda)\bar{\varphi}^{\rm LP}(\bar{\lambda}) & \varphi^{\rm LP}_f(\lambda)\varphi^{\rm LP}(-\lambda)-\varphi^{\rm LP}_f(-\lambda)\varphi^{\rm LP}(\lambda) \\
\bar{\varphi}^{\rm LP}_f(\bar{\lambda})\bar{\varphi}^{\rm LP}(-\bar{\lambda})-\bar{\varphi}^{\rm LP}_f(-\bar{\lambda})\bar{\varphi}^{\rm LP}(\bar{\lambda}) &
\bar{\varphi}^{\rm LP}_f(\bar{\lambda})\varphi^{\rm LP}(-\lambda)+\bar{\varphi}^{\rm LP}_f(-\bar{\lambda})\varphi^{\rm LP}(\lambda)
\end{array}\right) \nonumber
\end{eqnarray}
and an analogous expression for $V$.

Because $G_J$ is only a function of $k$, the LP matrices $U$ and $V$ calculated via \eref{LPMat} have no jump on $\Gamma_\Re$. Taking into account the absence of poles in $\vartheta^{IK}$, the LP matrices $U$ and $V$ can only have poles in $\pm\lambda_{1/2}$ as well as $\lambda=\infty$ and $\lambda=0$, respectively. They become holomorphic in $\mathbb{C}_{\lambda}\backslash\{\infty\}$ and $\mathbb{C}_{\lambda}\backslash\{0\}$ respectively, if we can arrange the $\varphi^{\rm LP}$-coefficients $p$, $q$ and $r$ so that poles at $\pm\lambda_{1/2}$ are prevented. Due to the symmetries of \eref{DchiU} it is again sufficient to investigate only the points $\lambda_{1/2}$. Thanks to $U_{22}=\bar{U}_{11}(\bar{\lambda})$ and $U_{21}=\bar{U}_{12}(\bar{\lambda})$ we only have to consider $U_{11}$ and $U_{12}$; the same applies to $V$.

We note that at $\lambda_{1/2}$ the exponents of $L_{1/2}$ and hence also the divergent behaviour  is preserved under the coordinate derivatives
\begin{equation}\begin{array}{ll}
L_{1f}=-\frac{\lambda\lambda_1}{f+g}L_1, &
  L_{2f}=-\frac{\lambda\lambda_2}{f+g}L_2,  \\
L_{1g}=-\frac{1}{\lambda\lambda_1(f+g)}L_1, &
  L_{2g}=-\frac{1}{\lambda\lambda_2(f+g)}L_2.
\end{array}\end{equation}

\subsection{Regularity condition for the LP matrix $U$ at $\lambda_2$}

The vanishing of the prefactor of $L_2$ in $(\det \Phi^{\rm LP}) U_{12}$ can be shown to be equivalent to
\begin{eqnarray}
\left(\kappa_{2f}+(2\rho_2-1)\frac{\lambda_2^2}{f+g}\kappa_2\right)(\psi_2^{\rm o}(\lambda_2))^2=0. \label{kappa2DGL}
\end{eqnarray}
For $\psi_2^{\rm o}(\lambda_2)\neq 0$ this leads together with the analogous calculation for $(\det \Phi^{\rm LP}) V_{12}$ to
\begin{eqnarray}
(\ln\kappa_2)_f=(1-2\rho_2)\frac{\lambda_2^2}{f+g}, \quad\quad
(\ln\kappa_2)_g=(1-2\rho_2)\frac{\lambda_2^{-2}}{f+g}. \label{kappa2DGL2}
\end{eqnarray}
The restriction to $\lambda=\lambda_2$ is enforced by setting $k=\frac12$ so that \eref{kappa2DGL2} can be read as differential equations for all $f$ and $g$. The system is integrable and has the solution
\begin{eqnarray}
\kappa_2=\rmi C_2\left(\frac{(\frac12+f)(\frac12-g)}{f+g}\right)^{2\rho_2-1},\quad C_2\in\mathbb{C}
\end{eqnarray}
where \eref{kappa2} yields even $C_2\in\mathbb{R}$. An exceptional solution to \eref{kappa2DGL} is given by $\psi_2^{\rm o}(\lambda_2)=0$ for all $f$ and $g$.

The vanishing of the prefactor of $L_2$ in $(\det \Phi^{\rm LP}) U_{11}$ at $\lambda_2$ can be shown to be equivalent to
\begin{eqnarray} \label{U22GL}
\fl\left[\kappa_{2f}+(2\rho_2-1)\frac{\lambda_2^2}{f+g}\kappa_2+(\kappa_2+\bar{\kappa}_2)\left(\frac{\psi_{2f}^{\rm o}(\lambda_2)}{\psi_{2}^{\rm o}(\lambda_2)}-\rho_2\frac{\lambda_2^2}{f+g}\right)\right]|\psi_2^{\rm o}(\lambda_2)|^2=0.
\end{eqnarray}
This equation is automatically fulfilled if \eref{kappa2} and \eref{kappa2DGL} hold. The same applies to the prefactor of $L_2$ in $(\det \Phi^{\rm LP}) V_{11}$.

\subsection{Regularity condition for the LP matrix $U$ at $\lambda_1$}

The evaluation of the LP matrix elements at $\lambda_1$ is similar, but we have to use additionally \eref{psipm} and its derivatives after identifying the $L_1$-prefactors and setting $k=-\frac12$. We get the analogous regularity conditions
for $(\det \Phi^{\rm LP}) U_{12}$ and $(\det \Phi^{\rm LP}) U_{11}$,
\begin{eqnarray}
\left(\kappa_{1f}+(2\rho_1-1)\frac{\lambda_1^2}{f+g}\kappa_1\right)(\psi_{1p}^{\rm o}(\lambda_1))^2=0, \label{kappa1DGL}\\ 
\fl\left[\kappa_{1f}+(2\rho_1-1)\frac{\lambda_1^2}{f+g}\kappa_1+(\kappa_1+\bar{\kappa}_1)\left(\frac{\psi_{1pf}^{\rm o}(\lambda_1)}{\psi_{1p}^{\rm o}(\lambda_1)}-\rho_1\frac{\lambda_1^2}{f+g}\right)\right]|\psi_{1p}^{\rm o}(\lambda_1)|^2=0 \label{U22GL2}
\end{eqnarray}
and equivalent relations for $(\det \Phi^{\rm LP}) V_{12}$ and $(\det \Phi^{\rm LP}) V_{11}$. Because the jump matrix $G_J$ depends only on $k$, the LP matrices $U$ and $V$ have no jump across $\Gamma_\Re$ and hence singularities associated with branch cuts are excluded as discussed in \sref{s:detPhiLambda1}. Again, \eref{U22GL2} is automatically fulfilled if \eref{kappa1} and \eref{kappa1DGL} hold. 

Together with their $V$-counterpart the combined regularity conditions at $\lambda_2$, \eref{kappa2}, \eref{kappa2DGL} and \eref{U22GL}, have the trivial solution $\psi_2^{\rm o}(\lambda_2)=0$ and similar the combined regularity conditions at $\lambda_1$, \eref{kappa1}, \eref{kappa1DGL} and \eref{U22GL2}, have the trivial solution $\psi_{1p}^{\rm o}(\lambda_1)=0$. The non-trivial solutions are
\begin{eqnarray}
\fl\kappa_2=\rmi C_2\left(\frac{(\frac12+f)(\frac12-g)}{f+g}\right)^{2\rho_2-1}
\hspace{-0.7cm},\hspace{.9cm}
\kappa_1=\rmi C_1\left(\frac{(\frac12-f)(\frac12+g)}{f+g}\right)^{2\rho_1-1}
\hspace{-0.7cm},\hspace{.9cm}
 C_1,C_2\in\mathbb{R}. \label{kappa12}
\end{eqnarray}
In each case 2 purely algebraic equations result for the 3 $\varphi^{\rm LP}$-coefficients $p$, $q$ and $r$, which have to be solved in step (iii) of the solution scheme in \fref{Lösungsschema}. Therefore a function of $f$ and $g$ may be left free to choose in the LP solution $\varphi^{\rm LP}$.



\subsection{Colliding wave conditions revisited}

In a last step (iv) of the solution scheme, the Ernst potential matching the initial data within the generated family of solutions to the LP  has to be identified. For this solution the colliding wave conditions \eref{cwc} are already fulfilled because of the appropriate choice of the initial data. Beyond this proper IVP solution, the family of solutions with the same jump matrix that results from the LP is also interesting. The fraction of the induced family of colliding wave spacetimes which obeys the colliding wave conditions generalises the proper IVP solution. 

Although the generic evaluation of the colliding wave conditions is beyond the scope of this article, we present a generic argument why the trivial solution $\psi_2^{\rm o}(\lambda_2)=0$ is supposed not to meet the colliding wave conditions and hence has to be excluded:
 For $f=\frac12$ the first partial contour $\Gamma_1$ vanishes and setting $\psi_2^{\rm o}(\lambda_2)=0$ the scalar LP-solution is due to \eref{psiugDarst} of the form 
\begin{equation}
\varphi_2^{\rm LP}=L_2^{\rho_2-1}\varphi_0^{\rm LP}, \quad \exists C<\infty: |\varphi_0^{\rm LP}|<C \;\forall \lambda.
\end{equation}
For $g=\frac12-\epsilon$, $\epsilon\ll 1$ the contour in the $k$-surface contracts to the twofold covering of $[\frac12-\epsilon,\frac12]$. There the difference $||J(k)-J(k=\frac12)||_\infty$ is bounded due to the Lipschitz continuity of $J(k)$. In the limit $g\to \frac12$ the deviation $||J(k)-J(k=\frac12)||_\infty$ of the jump matrix from its boundary value $J(k=\frac12)$ goes to zero, hence we conjecture $\varphi_2^{\rm LP}$ to have the same colliding wave limit as the solution $\varphi_{2c}^{\rm LP}=L_2^{\rho_2-1}$ for constant jump matrix $J(k)\equiv J(k=\frac12)$. Calculating $E_{2c}=\varphi_{2c}^{\rm LP}(1)/\varphi_{2c}^{\rm LP}(-1)$ we get 
\begin{equation}
\frac12\lim_{g\to\frac12}\left[\sqrt{\mbox{$\frac12$}-g}E_{2cg}(\mbox{$\frac12$},g)\right]=\rho_2-1. \label{Elim}
\end{equation}
This is not matching the initial values where we had defined $\rho_2:=|B_{\rm b}|$ and this is not even compatible with the allowed domain \eref{kWerte} of a colliding wave limit since we started with \eref{rhoGrenze2}. Therefore the trivial solution $\psi_2^{\rm o}(\lambda_2)=0$ and in the same way $\psi_{1p}^{\rm o}(\lambda_1)=0$ (where the corresponding scalar solutions $\varphi_1^{\rm LP}$ has to be examined in coordinates diagonalising $J(\pm\lambda_1)$) does not lead to correct colliding wave spacetimes and should be excluded. The non-trivial solutions in terms of this conjecture include a term behaving like $L_{1/2}^{\rho_{1/2}}$ at $\pm\lambda_{1/2}$. It should turn out to be dominant in some cases (at least for the proper IVP solution this is guaranteed) so that the limit corresponding to \eref{Elim} yields the right value $\rho_{1/2}$.

\section{Example: Generalisation of the Szekeres class of solutions}

\subsection{General solution of the linear problem}

In order to exemplify the solution generation technique embedded in the described inverse scattering procedure, we will study the generalisation of the Szekeres class of vacuum solutions \cite{Szekeres1972}. This class is a unification of the first exact colliding plane wave solutions including the Khan-Penrose solution and a step wave solution found even earlier by Szekeres. Remarkably, the Szekeres class of collinear polarised vacuum solutions also corresponds to a very easy solution in terms of the inverse scattering method. The scalar solution of the associated RHP is 
\begin{eqnarray}
\varphi_{Sz}=L_1^{\rho_1}L_2^{\rho_2}   \label{phiSz}
\end{eqnarray}
with the exponents $\rho_{1/2}$ varying in the range \eref{rhoGrenze} prescribed by the colliding wave conditions. The piecewise constant jump matrix is given by
\begin{eqnarray}
J=\left(\begin{array}{cc}
\alpha & 0 \\ 0 & \bar{\alpha}              
\end{array}\right)  \quad
{\rm with}\quad \alpha=\cases{ e^{-2\pi i\rho_2} & on $\Gamma_2$, \\
                               e^{2\pi i\rho_1}  & on $\Gamma_1$, \\
                                 1      & else.          }
\end{eqnarray}
Since it is diagonal everywhere on $\Gamma_\Re$, there is actually no need for the unitarisation transformation with $U^K$ and we also have $R_{\delta_1}=\mathbb{1}$.
However, in order to illustrate our procedure, we will literally stick to the full transformation formula \eref{cRHPjump} which leads to the cRHP jump matrix
\begin{eqnarray}
G_c^{IK}=\left(\begin{array}{cc} \alpha^{K} & 0 \\ 0 & \bar{\alpha}^{K} \end{array}\right),\quad
 \alpha^{K}=(w_+^K)^2 {\,\rm sign}^2(\Lambda_+^K)
\end{eqnarray}
and the scalar cRHP solutions 
\begin{eqnarray}
\vartheta^{IK}=w^K\Lambda^K.
\end{eqnarray} 
Via inverse transformation we get the 4 different scalar RHP solutions
\begin{eqnarray}
\varphi^{\rm{ee}}=L_1^{\rho_1-1}L_2^{\rho_2-1}, &
\varphi^{\rm{oe}}=L_1^{\rho_1-1}L_2^{\rho_2-1}(1+L_1), \\
\varphi^{\rm{eo}}=L_1^{\rho_1-1}L_2^{\rho_2-1}(1+L_2), \;\;&
\varphi^{\rm{oo}}=L_1^{\rho_1-1}L_2^{\rho_2-1}(1+L_1)(1+L_2).
\end{eqnarray}
The regularity conditions for the linear combination $\varphi^{\rm{LP}}=\varphi^{\rm{oo}}+p\varphi^{\rm{eo}}+q\varphi^{\rm{oe}}+r\varphi^{\rm{ee}}$ to satisfy the corresponding LP read 
\begin{eqnarray}
\kappa_2&=(\psi_{2}^{\rm o}(\lambda_2))^{-1} \left[\psi_{2}^{\rm o}(-\lambda_2)+q\psi_{2}^{\rm e}(-\lambda_2)\right]\overset{!}{=}\rmi C_2 L_{02}^{2\rho_2-1}, \\
\kappa_1&=(\psi_{1p}^{\rm o}(\lambda_1))^{-1} \left[\psi_{1p}^{\rm o}(-\lambda_1)+p\psi_{1p}^{\rm e}(-\lambda_1)\right]\overset{!}{=}\rmi C_1 L_{01}^{2\rho_1-1}
\end{eqnarray}
with
\begin{eqnarray}
\psi_2^{\rm e}=L_1^{\rho_1-1}\left[(1+L_1) +r/q\right], &
\psi_2^{\rm o}=L_1^{\rho_1-1}\left[(1+L_1) +p\right], \\
\psi_{1p}^{\rm e}=L_2^{\rho_2-1}\left[(1+L_2) +r/p\right],\;\; &
\psi_{1p}^{\rm o}=L_2^{\rho_2-1}\left[(1+L_2) +q\right] \label{psiHilf}
\end{eqnarray}
and $C_{1/2}\in\mathbb{R}$. Using the identity
\begin{eqnarray}
1+L_1L_2-L_{12}(L_1+L_2)=0 \label{LIdentität}
\end{eqnarray}
they can be evaluated to give the LP solution
\begin{eqnarray}
\fl\varphi^{\rm LP}
      &=L_1^{\rho_1} L_2^{\rho_2}\left[1-L_{12}^2+L_{12}^2\left(1+\rmi C_1 L_{01}^{2 \rho_1-1} L_{12}^{2 \rho_2-2}L_1^{-1}\right) \left(1+\rmi C_2 L_{02}^{2 \rho_2-1} L_{12}^{2\rho_1-2}L_2^{-1}\right)\right].  
\label{phiAllgSzek}
\end{eqnarray}
The corresponding Ernst potential $E=\varphi^{\rm LP}(1)/\varphi^{\rm LP}(-1)$ obeys the Ernst equation (which was already guaranteed by our procedure) and the colliding wave conditions without further restrictions. For this solution class the third functional degree of freedom not determined by the 2 regularity conditions turns out to be an overall factor in $\varphi$, which has already been omitted in \eref{phiAllgSzek} due to its insignificance for the physical Ernst potential. Nevertheless, we are left over with the 2 scalar real parameters $C_{1/2}$ in terms of which \eref{phiAllgSzek} is a generalisation of the Szekeres class \eref{phiSz}, which is reproduced for $C_1=0=C_2$. We note that also for the limiting case $\rho_{1/2}=\frac12$ of impulsive waves, which had been excluded in the derivation of our method, the expression \eref{phiAllgSzek} leads to a solution of the Ernst equation fulfilling the colliding wave conditions.

\subsection{Metric functions}

\begin{figure}[ht]
\centering
\includegraphics[width=0.45\linewidth]{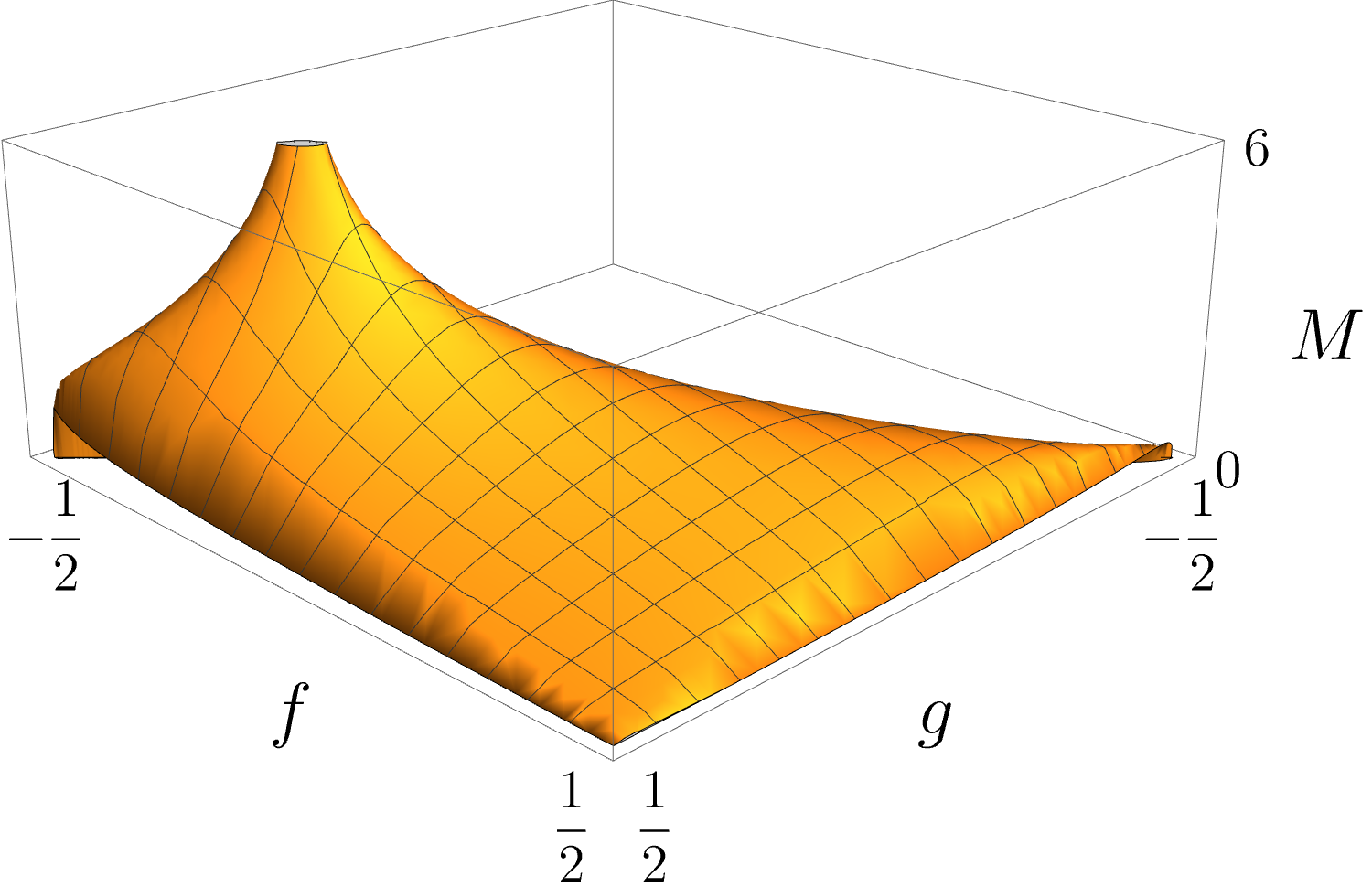}\qquad
\includegraphics[width=0.45\linewidth]{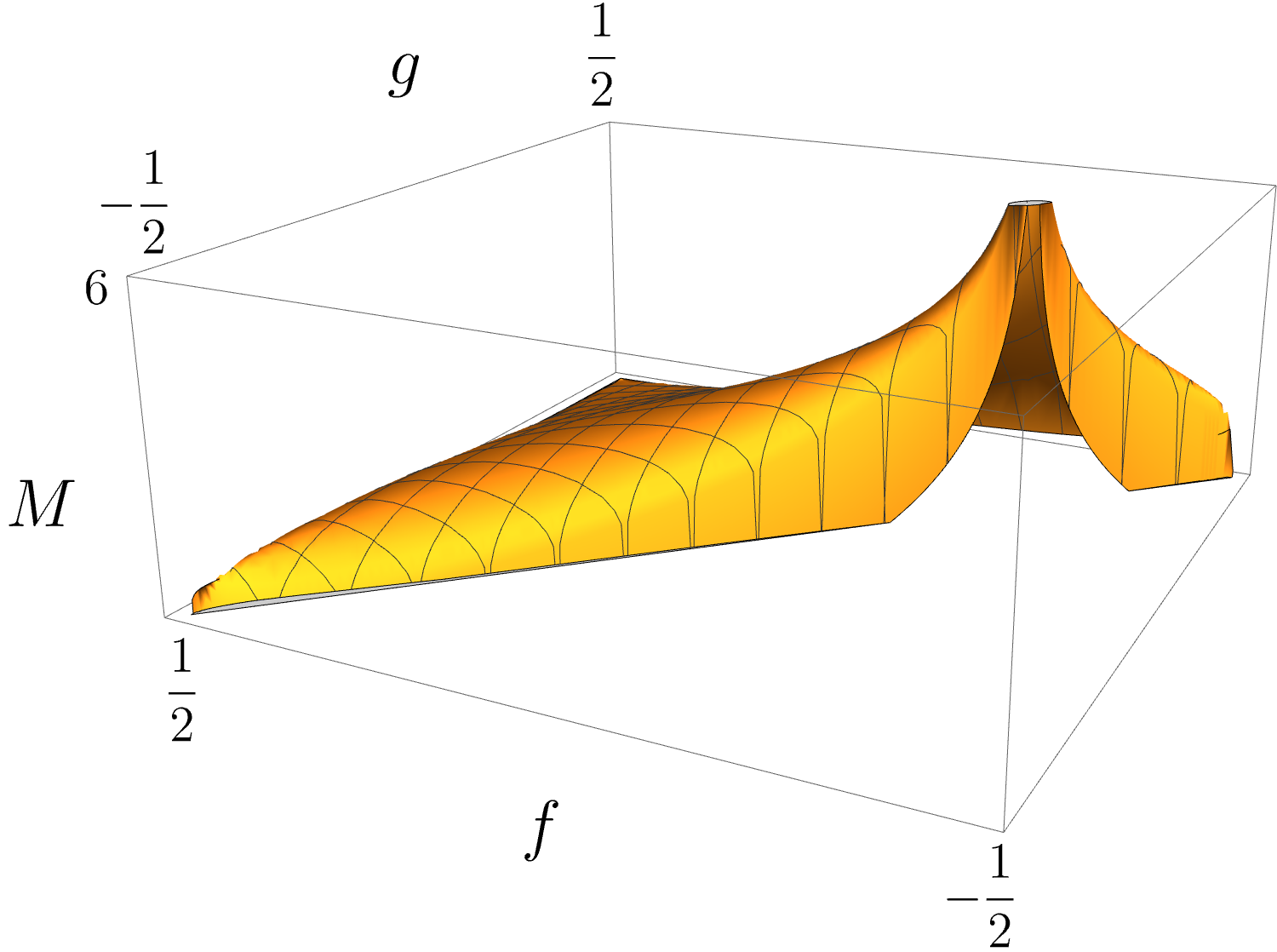}
\caption{The metric function $M$ in region IV viewed from two perspectives for $n_1=5$, $n_2=6$, $C_1=\frac15$, $C_2=\frac25$ featuring a pole at $\mathcal{D}=0$ on $f+g=0$.}
\label{fig:M}
\end{figure}

Using $L_{1/2p}:=L_{1/2}(1)=L_{1/2}^{-1}(-1)$ the Ernst potential $E=\varphi^{\rm LP}(1)/\varphi^{\rm LP}(-1)$ reads 
\begin{eqnarray}
\fl E&=L_{1p}^{2\rho_1}L_{2p}^{2\rho_2}\frac{1-L_{12}^2+L_{12}^2\left(1+\rmi C_1 L_{01}^{2 \rho_1-1} L_{12}^{2 \rho_2-2}L_{1p}^{-1}\right) \left(1+\rmi C_2 L_{02}^{2 \rho_2-1} L_{12}^{2\rho_1-2}L_{2p}^{-1}\right)}
{1-L_{12}^2+L_{12}^2\left(1+\rmi C_1 L_{01}^{2 \rho_1-1} L_{12}^{2 \rho_2-2}L_{1p}\right) \left(1+\rmi C_2 L_{02}^{2 \rho_2-1} L_{12}^{2\rho_1-2}L_{2p}\right)}. 
\label{EAllgSzek}
\end{eqnarray}
We note that for $C_{1/2}\neq 0$ the Ernst potential is complex and hence we generalised a class of collinear polarised waves to general polarisation.
Via the field equations \eref{Mu} and \eref{Mv} the last metric function $e^{-M}$ can be determined as
\begin{eqnarray}
\fl e^{-M}&=\frac{f_u g_v}{c_1 c_2 n_1 n_2 \sqrt{f+g}}  
  L_{01}^{-2\rho_1^2}L_{02}^{-2\rho_2^2}L_{12}^{-4\rho_1\rho_2}
  \left[\left(1+C_1 C_2 L_{01}^{2\rho_1-1}L_{02}^{2\rho_2-1}L_{12}^{2\rho_1+2\rho_2-2}\right)^2\right. \nonumber\\
\fl  &\hspace{7cm}\left.+\left(C_2 L_{02}^{2\rho_2-1}L_{12}^{2\rho_1}-C_1 L_{01}^{2\rho_1-1}L_{12}^{2\rho_2}\right)^2\right] \\
\fl  &=\frac{f_u g_v}{c_1 c_2 n_1 n_2 \sqrt{f+g}}  
  L_{01}^{-2\rho_1^2}L_{02}^{-2\rho_2^2}L_{12}^{-4\rho_1\rho_2}
  |\varphi_r(0)|^2
\label{eHochmMAllgSzek}
\end{eqnarray}
with the prefactor-reduced LP solution $\varphi_r:= L_1^{-\rho_1} L_2^{-\rho_2}\varphi^{\rm LP}$. The metric functions of the generalised Szekeres class have for $C_{1/2}\neq 0$ at the singularity $f+g=0$ a behaviour different from the Szekeres class, though it also leads to coordinate degeneracies: With $\epsilon=f+g>0$ we find for $\epsilon\to 0$ using $L_{1/2p}\sim\epsilon^{-1}$, $L_{01/2}\sim\epsilon^{-1}$, $L_{12}\sim\epsilon^{-1}$ the limits
\begin{eqnarray}
\fl E\sim-\rmi 2^{2\rho_1+2\rho_2-5} \mathcal{S},\;\;\;\; 
 \mathcal{S}:=2^{4\rho_1} (1-2f)^{3-2\rho_1-2\rho_2}C_1^{-1}+2^{4 \rho_2} (1+2f)^{3-2\rho_1-2\rho_2}C_2^{-1}, \\
\fl e^{-M}\sim\frac{16^{1+\rho_1^2+\rho_2^2}f_ug_v}{c_1c_2n_1n_2}\epsilon^{\frac12(2\rho_1+2\rho_2-3)(2\rho_1+2\rho_2-1)}(1-2f)^{-4\rho_1(\rho_1+\rho_2-1)}(1+2f)^{-4\rho_2(\rho_1+\rho_2-1)}\mathcal{D}^2, \nonumber \\
 \mathcal{D}:= 2^{-4\rho_1}(1-2f)^{2(\rho_1+\rho_2-1)}C_1 - 2^{-4\rho_2}(1+2f)^{2(\rho_1+\rho_2-1)}C_2.
\end{eqnarray}
In contrast to the divergence of the Szekeres class Ernst potential $E_{Sz}=L_{1p}^{2\rho_1}L_{2p}^{2\rho_2}$, in the general case the Ernst potential converges for all values of $\rho_{1/2}$ to a purely imaginary value at the singularity $f+g=0$ with a zero at $\mathcal{S}=0$ for opposite signs of $C_1$ and $C_2$. On the other hand, $e^{-M}$ diverges at $f+g=0$, whereas for the Szekeres class $e^{-M_{Sz}}$ vanishes. The inverse $e^{M}$ vanishes at $f+g=0$ with the exception of a pole at $\mathcal{D}=0$ for equal signs of $C_1$ and $C_2$, as can be studied in \fref{fig:M}. A physical interpretation of the relative sign between $C_1$ and $C_2$ will be given later. The plot of the Ernst potential in \fref{fig:E} shows a bump inside region IV.

\subsection{Weyl tensor components}

The scale invariant Weyl tensor components $\Psi_i^{\circ}$ (cf. \cite{Griffiths1991}) can be represented as
\begin{eqnarray}
\Psi_0^{\circ}&=g_v^2\frac{{\rm sign}^{-1}\left[\varphi_r(1)\varphi_r(-1)\varphi_r^2(\infty)\right]}
   {2(f+g)^2\varphi_r(\infty)}P_0, \\
\Psi_4^{\circ}&=f_u^2\frac{{\rm sign}\left[\varphi_r(1)\varphi_r(-1)\varphi_r^2(0)\right]}
   {2(f+g)^2\varphi_r(\infty)}P_4, \\
\Psi_2^{\circ}&=f_ug_v\frac{{\rm sign}^{-1}\left[\varphi_r^2(\infty)\right]}
   {4(f+g)^2\varphi_r(\infty)}P_2,
\end{eqnarray}
using the expressions 
\begin{eqnarray}
\fl P_0:=F_0(\rho_1,\rho_2)+C_1 C_2 L_{01}^{2 \rho_1-1} L_{02}^{2 \rho_2-1} L_{12}^{2 \rho_1+2\rho_2-2} F_0(\rho_1-1,\rho_2-1)\\
\fl   \qquad\qquad\qquad\qquad -\rmi C_1 L_{01}^{2 \rho_1-1} L_{12}^{2
   \rho_2} F_0(\rho_1-1,\rho_2)+\rmi C_2 L_{02}^{2 \rho_2-1} L_{12}^{2
   \rho_1} F_0(\rho_1,\rho_2-1), \nonumber\\
\fl P_4:=F_4(\rho_1,\rho_2)+C_1 C_2 L_{01}^{2 \rho_1-1} L_{02}^{2 \rho_2-1} L_{12}^{2 \rho_1+2\rho_2-2} F_4(\rho_1-1,\rho_2-1)\\
\fl   \qquad\qquad\qquad\qquad -\rmi C_1 L_{01}^{2 \rho_1-1} L_{12}^{2
   \rho_2} F_4(\rho_1-1,\rho_2)+\rmi C_2 L_{02}^{2 \rho_2-1} L_{12}^{2
   \rho_1} F_4(\rho_1,\rho_2-1), \nonumber\\
\fl P_2:=F_2(\rho_1,\rho_2)+C_1 C_2 L_{01}^{2 \rho_1-1} L_{02}^{2 \rho_2-1} L_{12}^{2 \rho_1+2\rho_2-2} F_2(\rho_1-1,\rho_2-1)\\
\fl   \qquad\qquad\qquad\qquad +\rmi C_1 L_{01}^{2 \rho_1-1} L_{12}^{2
   \rho_2} F_2(\rho_1-1,\rho_2)-\rmi C_2 L_{02}^{2 \rho_2-1} L_{12}^{2
   \rho_1} F_2(\rho_1,\rho_2-1), \nonumber\\
F_0(a,b):=4(a\lambda_1^{-1}+b\lambda_2^{-1})^3-a\lambda_1^{-3}-b\lambda_2^{-3}, \\
F_4(a,b):=4(a\lambda_1+b\lambda_2)^3-a\lambda_1^{3}-b\lambda_2^{3}, \\
F_2(a,b):=4ab\lambda_1^{-1}\lambda_2^{-1}(\lambda_1-\lambda_2)^2+4(a+b)^2-1.
\end{eqnarray}
The $\Psi_i^{\circ}$ are invariant under a rescaling of the lightlike null tetrad vectors. $\Psi_4^{\circ}$ is the only non-vanishing component for the left initial wave in region II and only $\Psi_0^{\circ}$ is non-vanishing for the right wave in region III. The so-called `Coulomb component' $\Psi_2^{\circ}$ arises in region IV due to the nonlinear interaction of these incoming waves. 
A transition to the Weyl tensor components w.r.t. the symmetric lightlike tetrad vectors $l_a=\rme^{-\frac12 M}\delta^u_a$, $n_a=\rme^{-\frac12 M}\delta^v_a$ is achieved via $\Psi_0=e^{M}\Psi_0^{\circ}$ and $\Psi_4=e^{M}\Psi_4^{\circ}$. These components are suitable for a discussion of the incoming waves, although they are in principle coordinate dependent as well as tetrad dependent. A `wave profile' with an invariant meaning could be calculated by transformation to the Brinkmann form of the metric, which is out of the scope of this article. With $R_1:=\rmi C_1\left(\frac{1-2f}{1+2f}\right)^{2\rho_1-1}$ and $R_2:=\rmi C_2\left(\frac{1-2g}{1+2g}\right)^{2\rho_2-1}$ the incoming waves can be described by
\begin{eqnarray}
\fl\left.\Psi_0\right\vert_{f=\frac12}=\frac12 c_2^2 n_2^2\frac{(\frac12-g)^{4\rho_2^2-3/2}(2\rho_2-1)\left[\rho_2(1+2\rho_2)+R_2(3-5\rho_2+2\rho_2^2)\right]}
   {(\frac12+g)^{2\rho_2^2+3/2}(1+R_2)^3{\rm \;sign}\left[(1+2g)(1-R_2)^2+8R_2\right]}, \\
\fl\left.\bar{\Psi}_4\right\vert_{g=\frac12}=\frac12 c_1^2 n_1^2\frac{(\frac12-f)^{4\rho_1^2-3/2}(2\rho_1-1)\left[\rho_1(1+2\rho_1)+R_1(3-5\rho_1+2\rho_1^2)\right]}
   {(\frac12+f)^{2\rho_1^2+3/2}(1+R_1)^3{\rm \;sign}\left[(1+2f)(1-R_1)^2+8R_1\right]} 
\end{eqnarray}
using a perfect analogy between $\Psi_0$ and $\bar{\Psi}_4$. Near the wave front $f=\frac12$ we have 
\begin{eqnarray}
\left.\bar{\Psi}_4\right\vert_{g=\frac12}\sim \mbox{$\frac12$} c_1^2n_1^2\rho_1(4\rho_1^2-1)(\mbox{$\frac12$}-f)^{-3/2+4\rho_1^2},
\end{eqnarray}
which is the same asymptotical behavoiur as for the Szekeres class. For $\frac12\leq\rho_1<\sqrt{3/8}$, i.e. $2\leq n_1<4$, the incoming Weyl tensor component $\left.\bar{\Psi}_4\right\vert_{g=\frac12}$ is unbounded at the wave front $f=\frac12$; for $\sqrt{3/8}\leq\rho_1<\sqrt{1/2}$ ($4\leq n_1$) it is bounded.

At the fold singularity $f=-\frac12$ the Weyl tensor component diverges as 
\begin{eqnarray}
\left.\bar{\Psi}_4\right\vert_{g=\frac12}\sim \rmi c_1^2 n_1^2\frac{3-11\rho_1+12\rho_1^2-4\rho_1^3}{2C_1^2{\rm \;sign}(C_1)}(\mbox{$\frac12$}+f)^{-3/2-2(1-\rho_1)^2}.
\end{eqnarray}
with purely imaginary coefficient, whereas the Szekeres class had the stronger divergence behaviour
\begin{eqnarray}
\left.\bar{\Psi}_4\right\vert_{g=\frac12}\sim \frac12 c_1^2 n_1^2\rho_1(4\rho_1^2-1)(\mbox{$\frac12$}+f)^{-3/2-2\rho_1^2}.
\end{eqnarray}
This divergence is a strong hint for the existence of a non-scalar curvature singularity at the boundary $f=-\frac12,v<0$ of region II. This singularity character has already been confirmed for the Szekeres class  \cite{Helliwell_1989}.

\subsection{The limit of circularly polarised impulsive waves}

\begin{figure}[ht]
\centering
\includegraphics[width=0.49\linewidth]{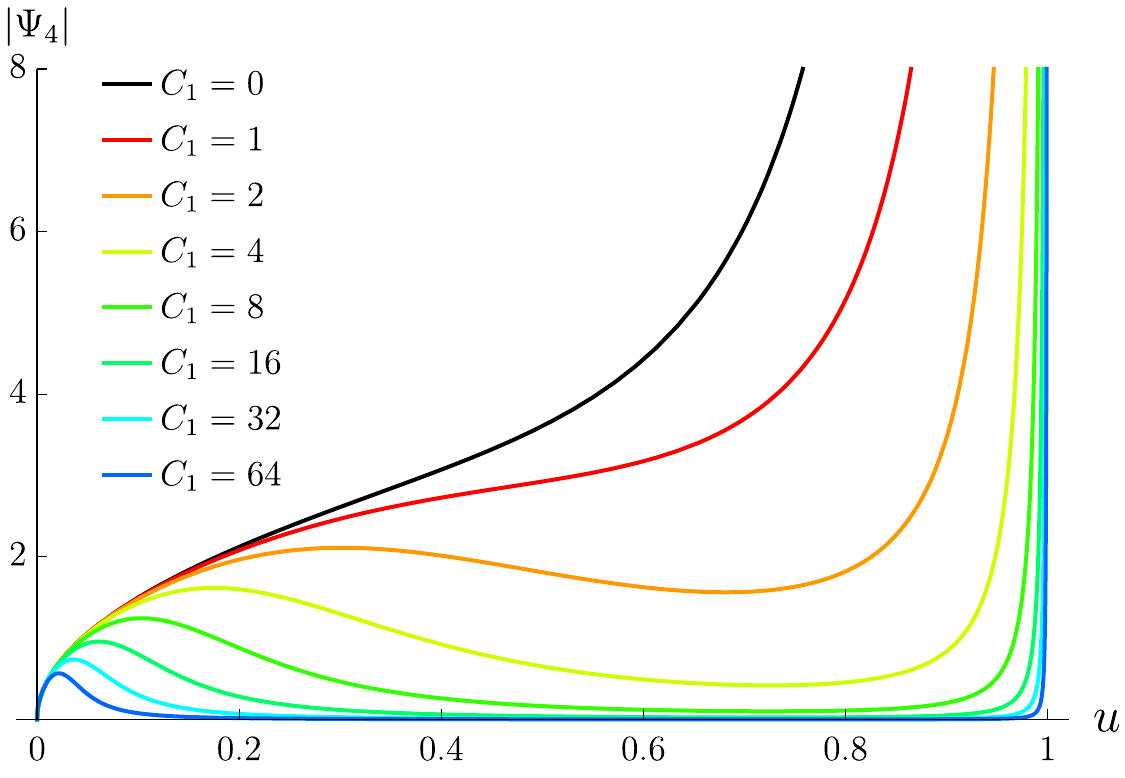}
\includegraphics[width=0.49\linewidth]{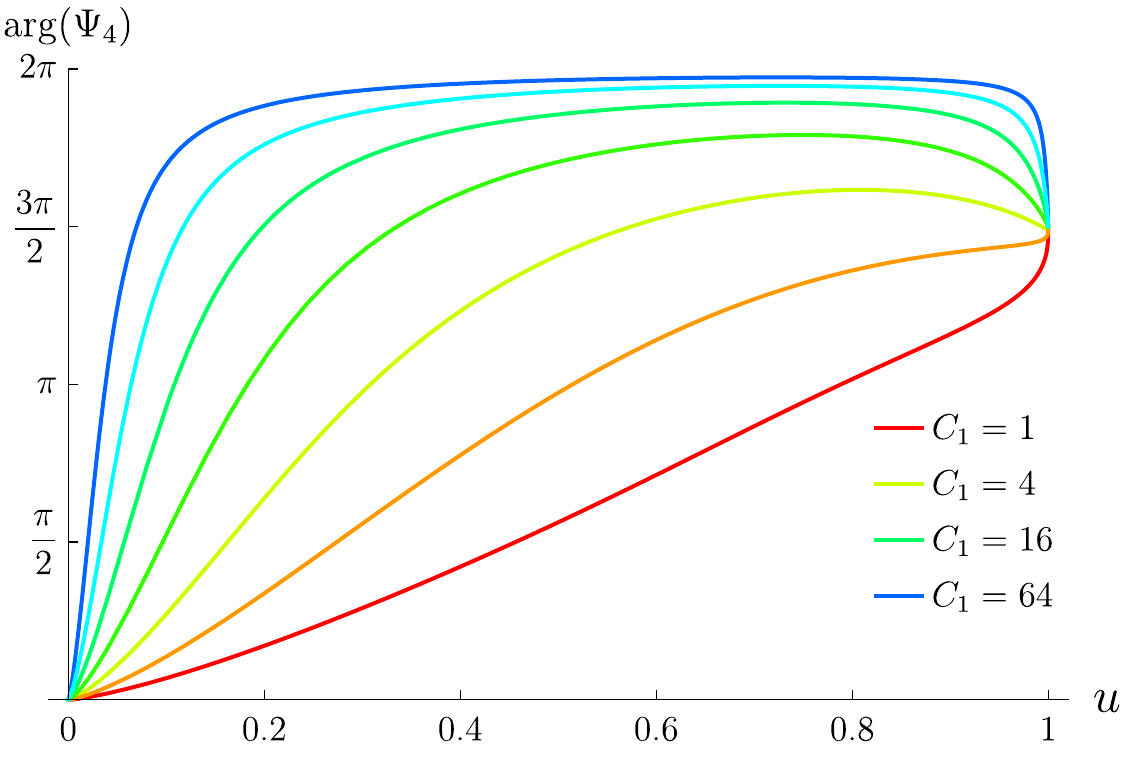}
\caption{Amplitude and polarisation of the initial Weyl tensor component $\left.\Psi_4\right\vert_{g=\frac12}$ for $n_1=5$ and $c_1=1$ approximating a circularly polarised pulsed wave for increasing $C_1$.}
\label{fig:Initial_profile}
\end{figure}

Another interesting aspect of this generalisation of the Szekeres class is the shape of the Weyl tensor components for large values of $C_{1/2}$. As illustrated in \fref{fig:Initial_profile}, $\left.\Psi_4\right\vert_{g=\frac12}$ compactifies into a pulse at the wave front $f=\frac12$ ($u=0$) for increasing $C_1$ featuring a full revolution of the polarisation angle during that pulse. Note that the `wave strength' $c_1$ has been fixed to $1$ in \fref{fig:Initial_profile}, but can be easily modified to adjust the height of the pulse. In consequence, this generalised solution class can provide analytical formulas for a new type of circularly polarised impulsive waves. However, some attention may have to be paid to the leftover divergence of $\left.\Psi_4\right\vert_{g=\frac12}$ at the fold singularity $f=-\frac12$ ($u=1$ for $c_1=1$).

As \fref{fig:Initial_profile} indicates, the signs of $C_1$ and $C_2$ describe the direction of rotation in the incoming waves' phases: The zero in the Ernst potential $E$ related to opposite signs occures for opposite rotational directions of $\left.\bar{\Psi}_4\right\vert_{g=\frac12}$ and $\left.\Psi_0\right\vert_{f=\frac12}$, the zero in the $e^{-M}$ related to equal signs occures for equal rotational directions of $\left.\bar{\Psi}_4\right\vert_{g=\frac12}$ and $\left.\Psi_0\right\vert_{f=\frac12}$. 

\subsection{The character of the singularity at $f+g=0$}

Finally we compute the first scalar curvature invariant of the Weyl tensor for the generalised Szekeres class,
\begin{eqnarray}
\fl \mathcal{I}&=16e^{2M}(3(\Psi_2^{\circ})^2+\Psi_0^{\circ}\Psi_4^{\circ}) 
=\frac{c_1^2 c_2^2 n_1^2 n_2^2 L_{01}^{4 \rho_1^2}L_{02}^{4 \rho_2^2} L_{12}^{8 \rho_1 \rho_2}}{(f+g)^3\varphi_{r}^6(\infty)}(3P_2^2+4P_0P_4).
\end{eqnarray}
At $\epsilon=f+g\to 0$ it diverges for $C_{1,2}\neq 0$ like
\begin{eqnarray}
\fl\mathcal{I}\sim c_1^2 c_2^2 n_1^2 n_2^2 \frac{(1-2f)^{8\rho_1(\rho_1+\rho_2-1)}(1+2f)^{8\rho_2(\rho_1+\rho_2-1)}}{4^{1+\rho_1^2+\rho_2^2}\mathcal{D}^4}\epsilon^{-3-4(\rho_1+\rho_2-1)^2}P_{\mathcal{I}}, \\
P_{\mathcal{I}}:=3 \left[1-4 (\rho_1+\rho_2-1)^2\right]^2+4 \left[4 (\rho_1+\rho_2-1)^3-\rho_1-\rho_2+1\right]^2.
\end{eqnarray}
Therefore the boundary $f+g$ of region IV is a scalar curvature singularity as it is for the Szekeres class. Nevertheless, the divergence is weaker than for the Szekeres class where the curvature invariant behaves like
\begin{eqnarray}
\fl\mathcal{I}_{Sz}\sim c_1^2 c_2^2 n_1^2 n_2^2\frac{(1-2f)^{8\rho_1(\rho_1+\rho_2)}(1+2f)^{8\rho_2(\rho_1+\rho_2)}}{4^{4(\rho_1^2+\rho_2^2)}\epsilon^{3+4(\rho_1+\rho_2)^2}}\left[1-4(\rho_1+\rho_2)^2\right]^2\left[3+4(\rho_1+\rho_2)^2\right].\nonumber
\end{eqnarray}
In case of equal rotational directions of $\left.\bar{\Psi}_4\right\vert_{g=\frac12}$ and $\left.\Psi_0\right\vert_{f=\frac12}$ there is a pole structure at $\mathcal{D}=0$ on top of the divergence behaviour at the boundary $f+g=0$, as can be seen in \fref{fig:I}. The exact position of that pole is determined by the ratio of $C_1$ and $C_2$.

\section{Conclusions}

With the inverse scattering method and the subsequent transformation to a cRHP we were able to construct a solution to the characteristic initial value problem of colliding plane waves. 

For a given set of initial values the crucial problem consists in the solution to the integral equation belonging to \eref{Sprungtheta}, whereas the derivation of the jump matrix from the initial data via the ODE \eref{alphabeta1}-\eref{ODE2} is possible with high numerical accuracy, if not analytically. Although the jump matrix can be only approximated numerically for generic initial data, the transformation to the cRHP only depending on $J(\pm\lambda_{1/2})$ is given analytically by \eref{cRHPsolution} and \eref{cRHPjump}. The regularity conditions \eref{kappa12} adapting the RHP solution to the LP are algebraic and finally left over degrees of freedom have to be fixed by comparison with the initial data. 

In special cases where a fully analytic treatment is possible, the fourfold ambiguity contained in the solution to the discontinuous RHP and the possible remnant functional degree of freedom in the LP solution leads to the construction of families of exact solutions. In this sense the described procedure serves as a solution generating technique which generalises existing colliding wave solutions and leads to insights into the structure of colliding plane waves. This was demonstrated by generalisation of the collinear polarised Szekeres class of colliding wave spacetimes to a class with general polarisation. A scalar curvature singularity in the interaction region has been identified for this class and evidence for a non-scalar curvature singularity at the `fold singularity' has been given. Moreover, a possible limiting case with circularly polarised impulsive waves has been discovered. A more rigorous generic treatment of the colliding wave conditions for the family of spacetimes induced by the LP solutions is subject of ongoing investigations.

For an impulsive wave the boundary value of the corresponding RHP jump matrix takes the value $J(\pm\lambda_i)=-\mathbb{1}$ ($i=1$ for an impulsive wave in region II, $i=2$ for an impulsive wave in region III) which is invariant under rotation transformation and unitarisation transformation. Hence these types of transformation can be used to set the derivatives of the jump functions $\beta$ and $\gamma$ to zero at $\pm\lambda_i$ instead of their values. After  appropriate preparation the discontinuities in the eRHP can be removed by the alternative singularity transformations $S_{i}^{\rm e}$ and $(S_{i}^{\rm e})^{-1}$ instead of $S_{i}^{\rm e}$ and $S_{i}^{\rm o}$.  The inverse transformation leads directly (i.e. without linear combinations) to the construction of four RHP solutions out of the sRHP solutions. Since $\rho_i=1-\rho_i=\frac12$ the derivation of the regularity conditions has to be recapitulated carefully for a spacetime with at least one impulsive wave, but we expect a simplification in the end. Massive simplifications in the described solution procedure occur also for  initially collinearly polarised waves. 

\ack
This research was supported by the Konrad-Adenauer-Stiftung and the Deutsche Forschungsgemeinschaft (DFG) through the Graduiertenkolleg 1532 `Quantum and Gravitational Fields'. We thank David Hilditch for useful discussions.

\newpage

\appendix

\section{Visualisation of the transformation of the jump functions}

\begin{figure}[ht]
\centering
\includegraphics[width=0.48\linewidth]{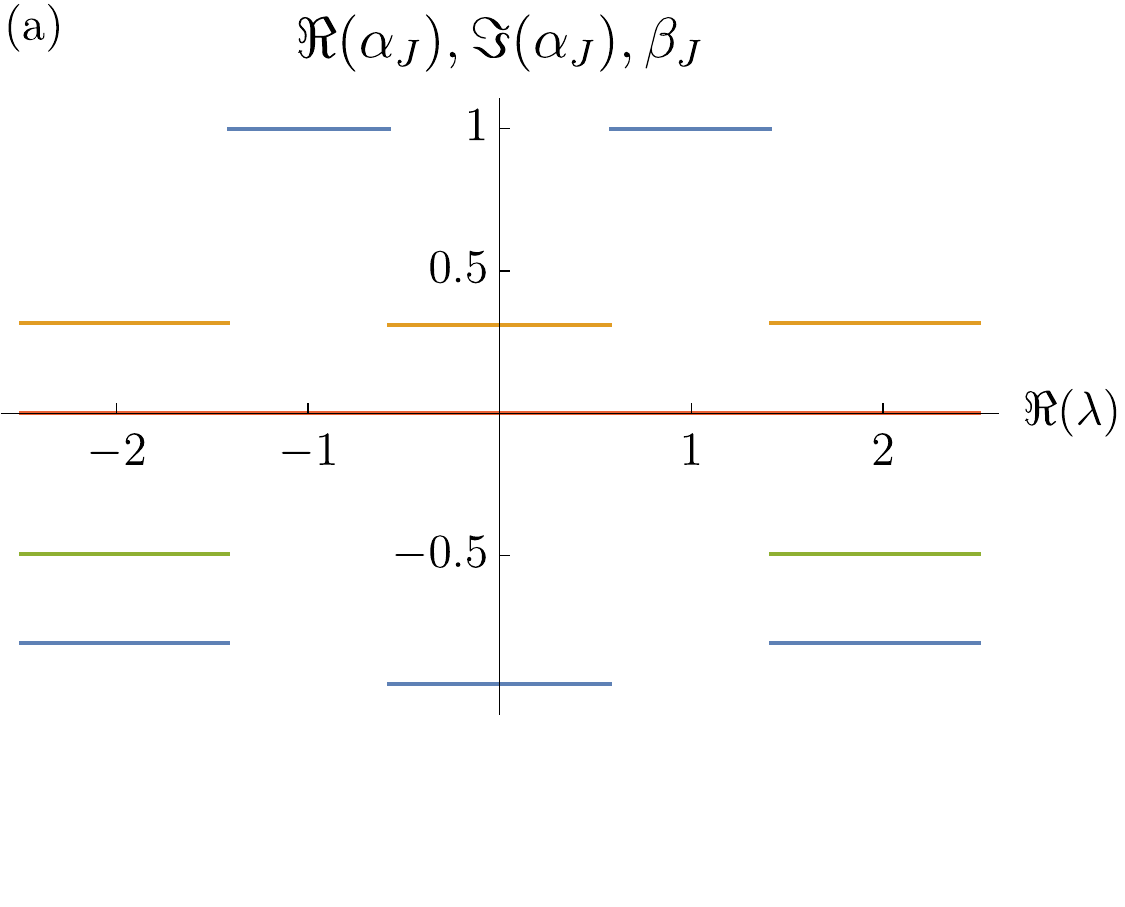}\;\;
\includegraphics[width=0.48\linewidth]{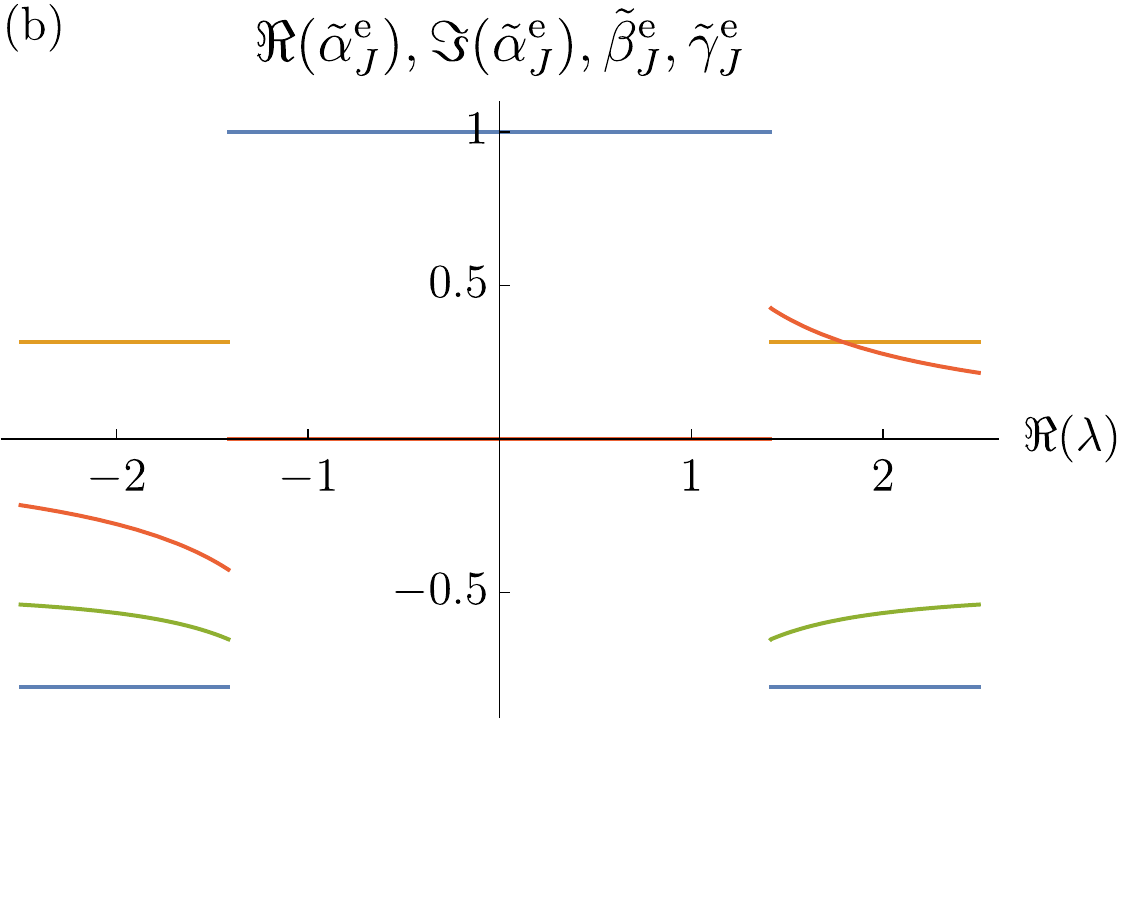}
\includegraphics[width=0.48\linewidth]{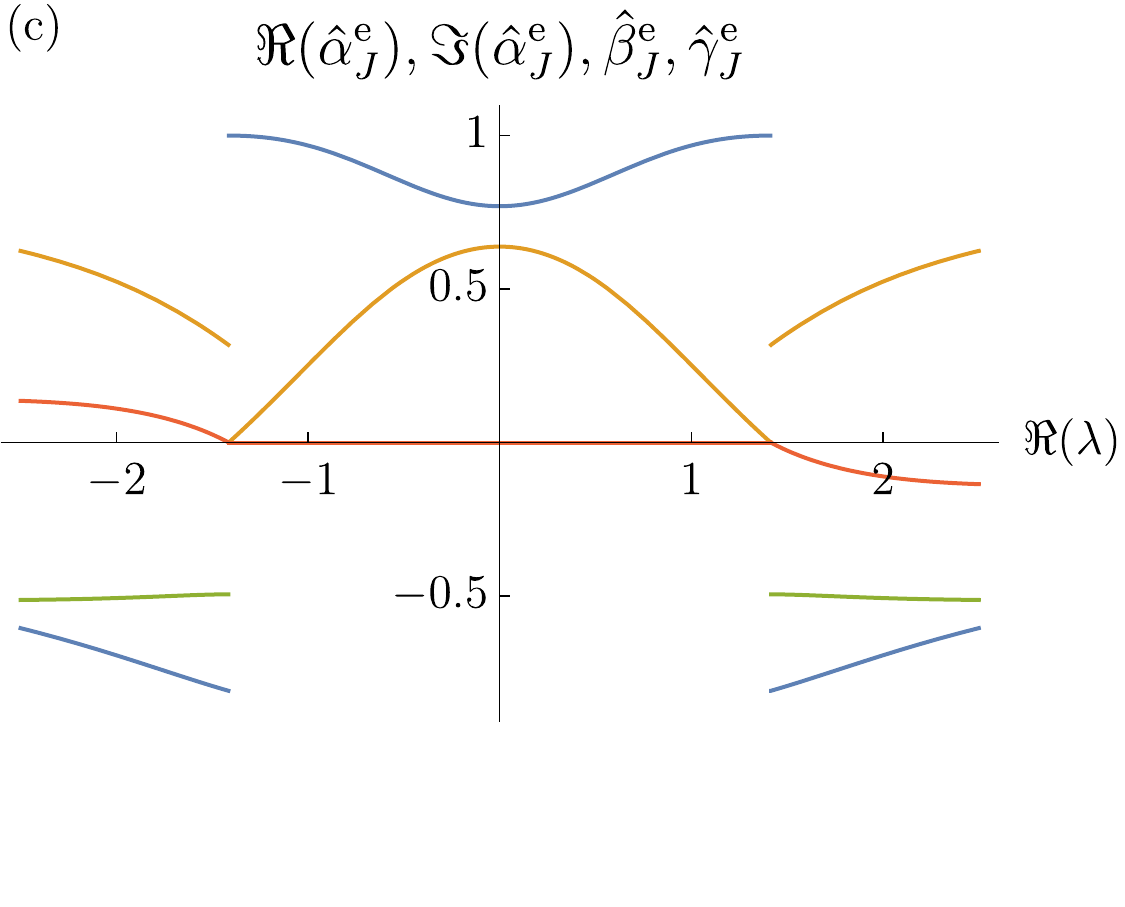}\;\;
\includegraphics[width=0.48\linewidth]{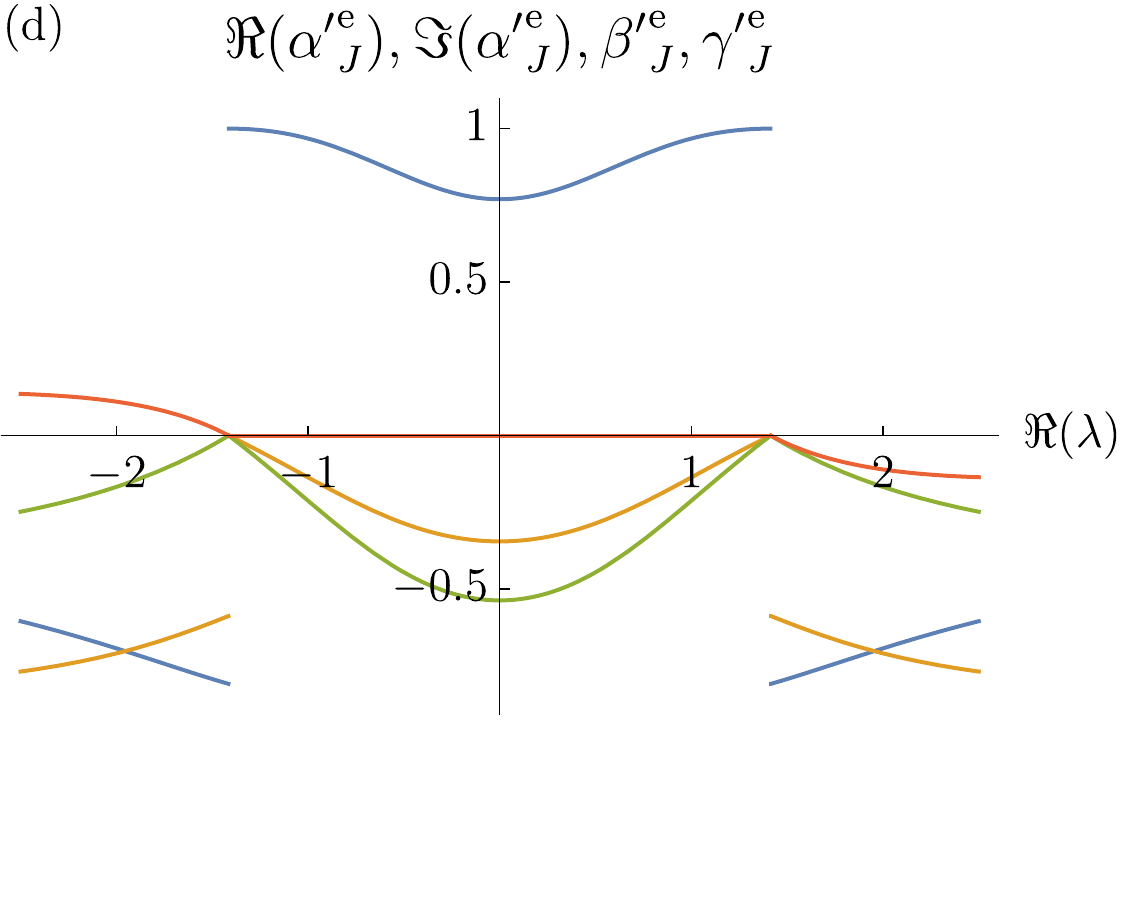}
\includegraphics[width=0.60\linewidth]{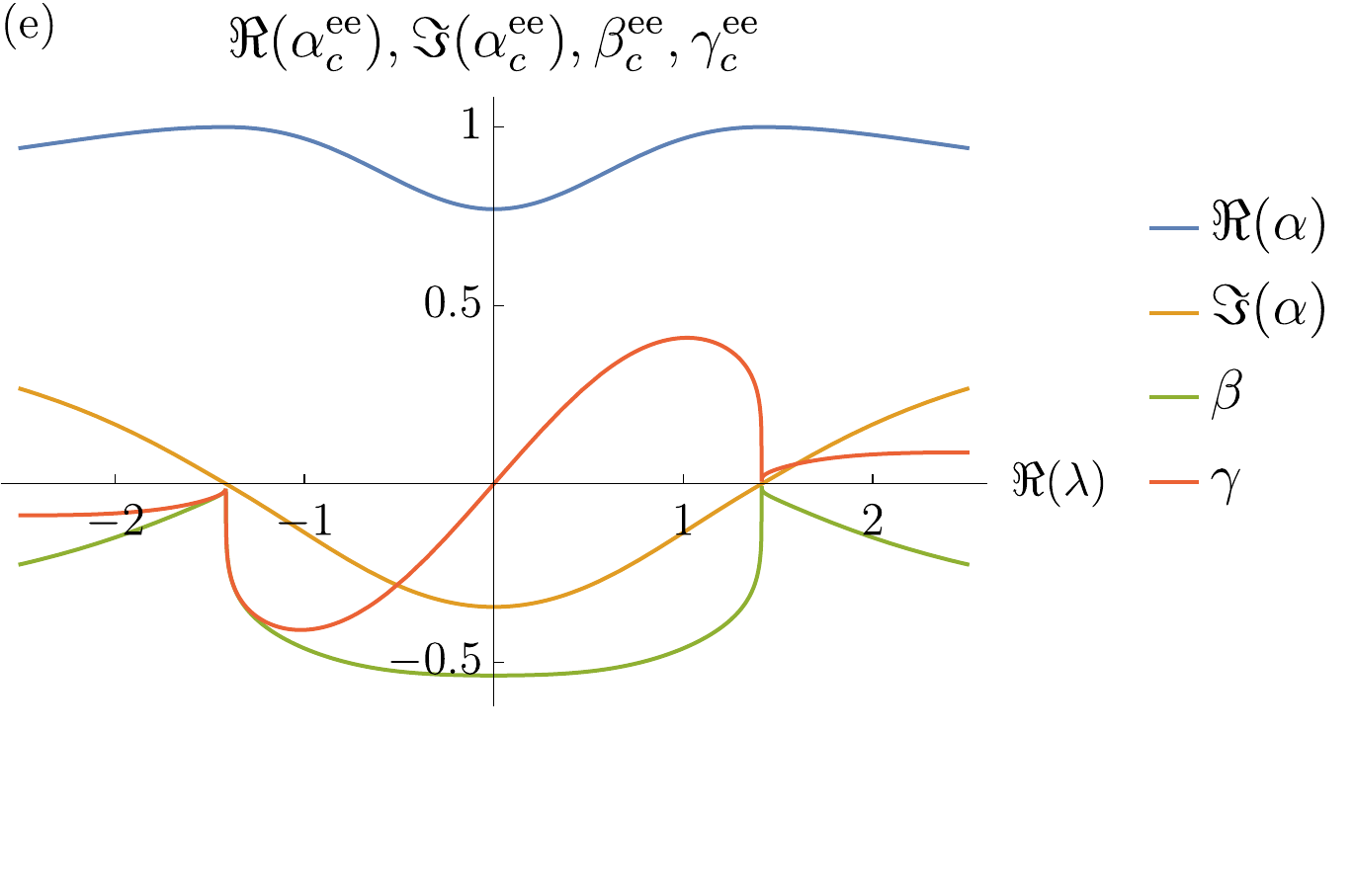}
\caption{Visualisation of the transformation of the jump functions $\Re(\alpha)$ (blue), $\Im(\alpha)$ (orange), $\beta$ (green) and $\gamma$ (red) contained in the jump matrices $G_J$ (a), $\tilde{G}_J^{\rm e}=(S^e_{2-})^{-1} G_J S^e_{2+}$ (b), $\hat{G}_J^e=(U^{e}_-)^{-1}\tilde{G}_J^eU^e_+$ (c), ${G'}_J^e=R_{\delta_1}^{-1}\hat{G}_J^eR_{\delta_1}$ (d) and $G_c^{\rm ee}=(S^e_{1-})^{-1} {G'}_J^e S^e_{1+}$ (e). The initial jump matrix is chosen piecewise constant and diagonal on $\Gamma_2$.}
\label{fig:VisTrafo}
\end{figure} 

\section{Visualisation of the generalised Szekeres class}

\begin{figure}[ht]
\centering
\includegraphics[width=0.45\linewidth]{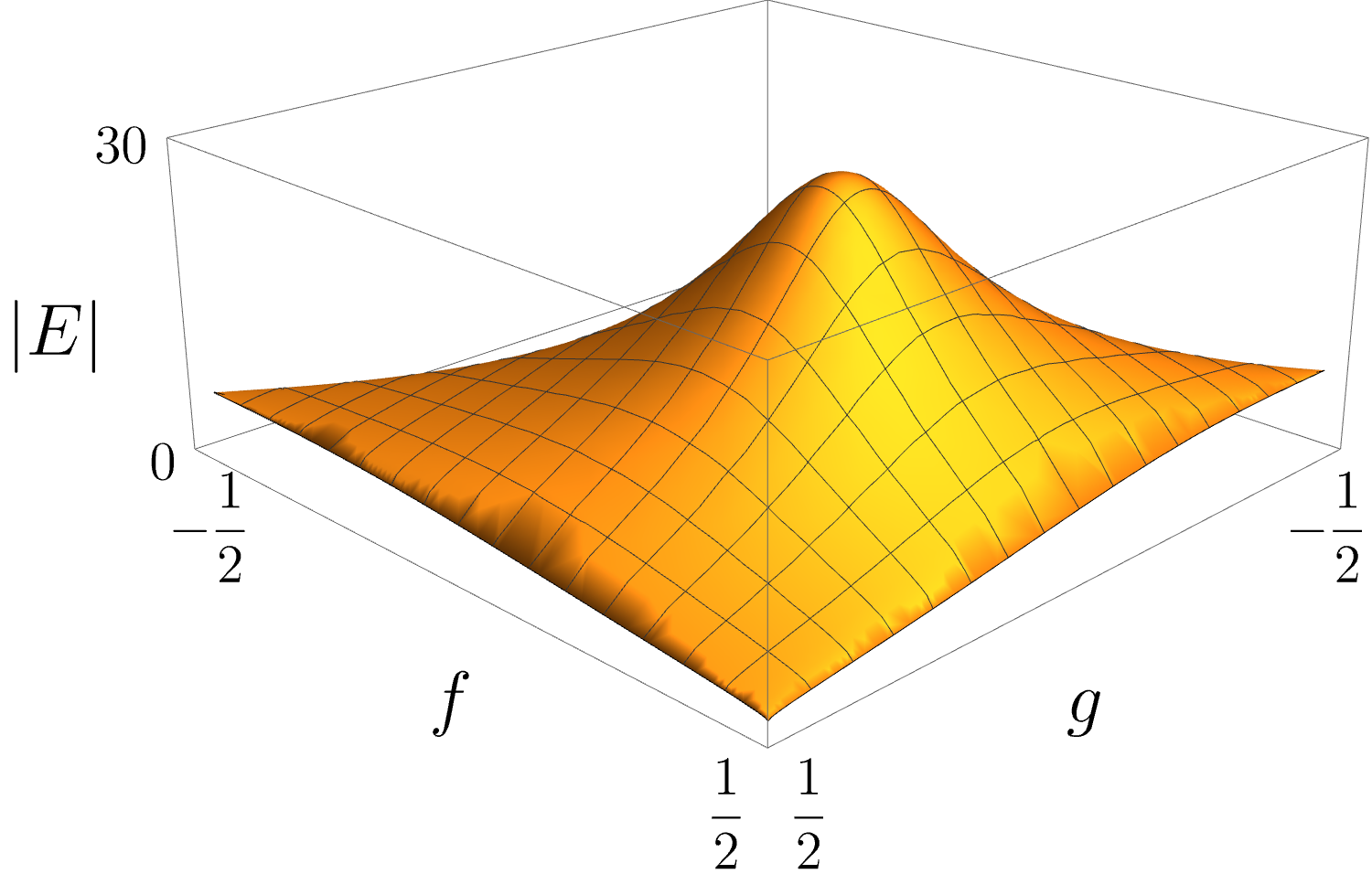}\qquad
\includegraphics[width=0.45\linewidth]{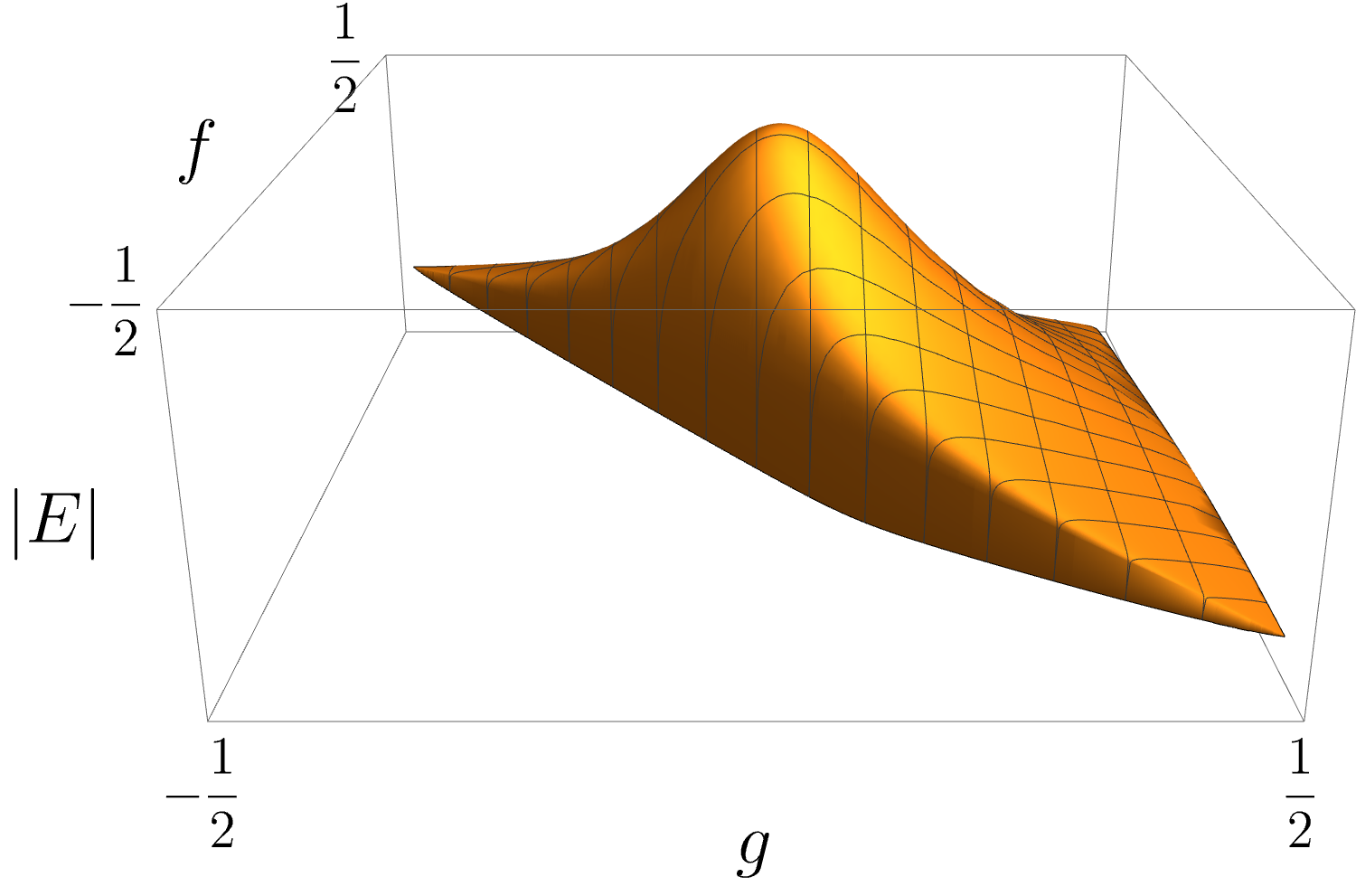}
\caption{The absolute value of the Ernst potential $E$ in region IV viewed from two perspectives for $\rho_1=\frac{55}{100}$, $\rho_2=\frac{6}{10}$, $C_1=\frac15$, $C_2=-\frac16$ featuring a zero at $\mathcal{S}=0$ on $f+g=0$ and a bump inside region IV.}
\label{fig:E}
\end{figure} 

\begin{figure}[ht]
\centering
\includegraphics[width=0.45\linewidth]{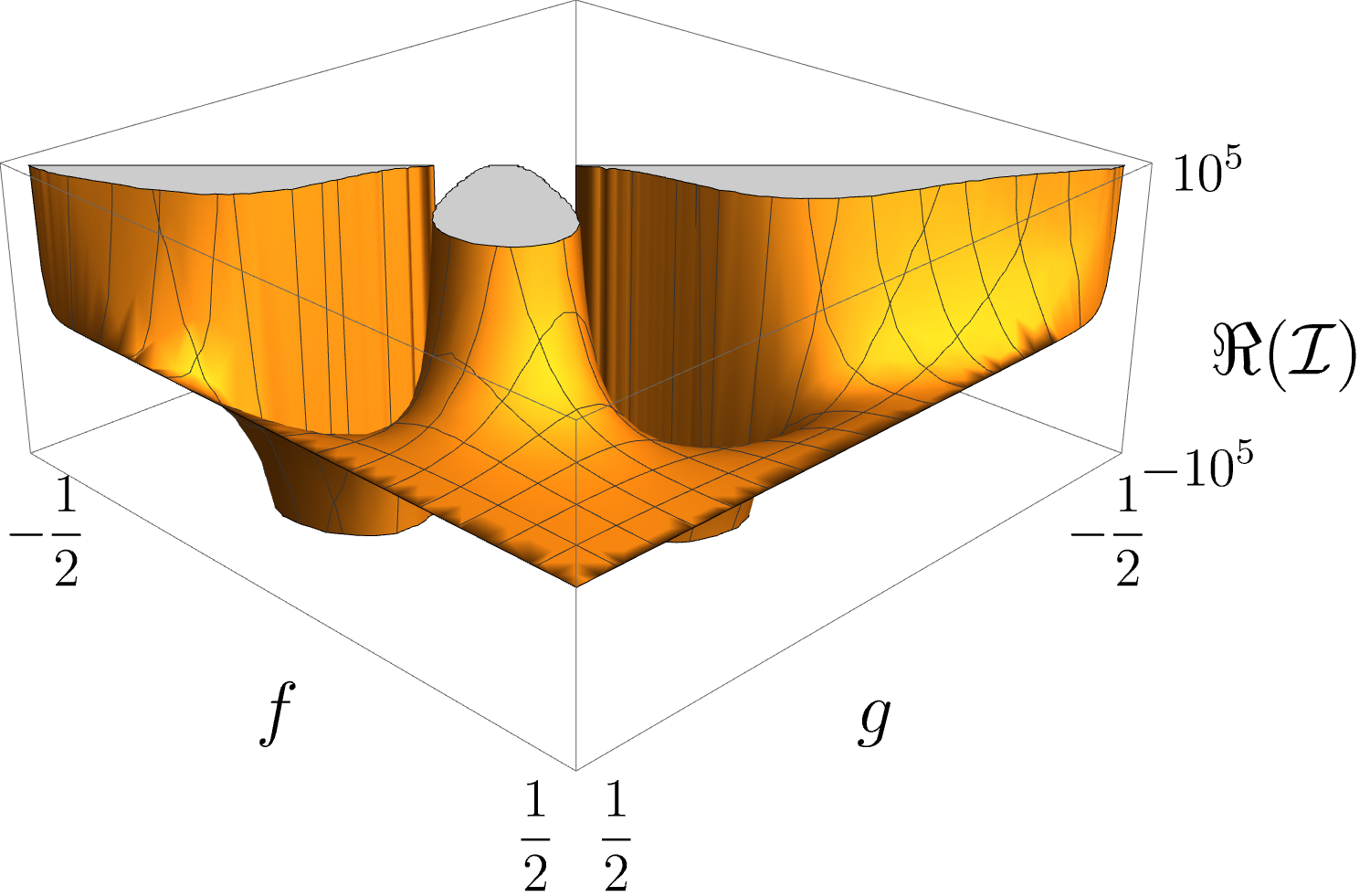}\qquad
\includegraphics[width=0.45\linewidth]{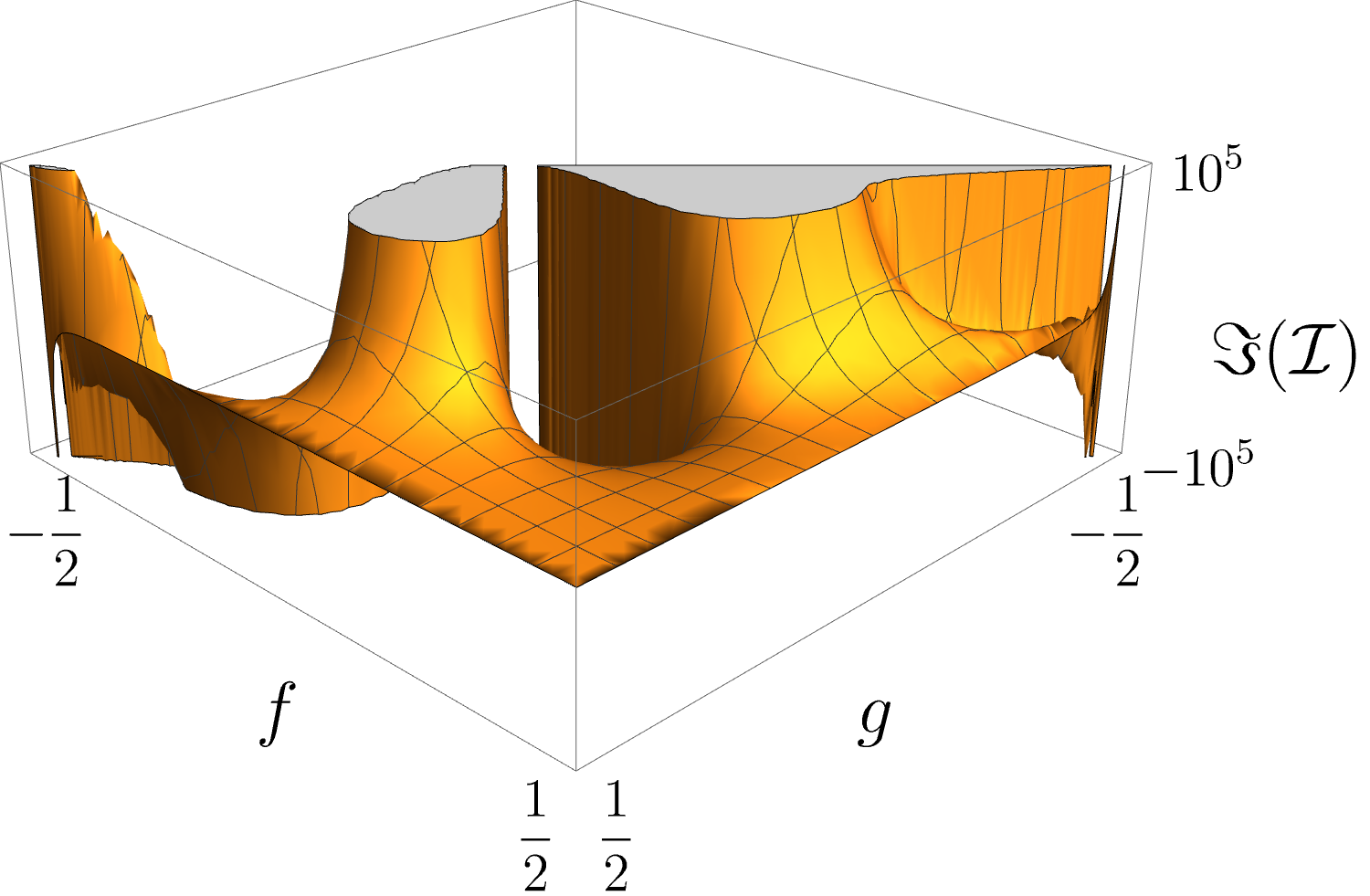}
\caption{Real and imaginary part of the scalar invariant $\mathcal{I}$ in region IV  for $n_1=5$, $n_2=6$, $C_1=1$, $C_2=\frac65$ featuring a higher order pole at $\mathcal{D}=0$ on $f+g=0$.}
\label{fig:I}
\end{figure}

\end{document}